\documentclass[preprint,11pt,authoryear]{elsarticle}

\usepackage[a4paper, top=2.5cm, bottom=2.5cm, left=2.5cm, right=2.5cm]{geometry}
\usepackage{setspace}
\usepackage{amssymb}
\usepackage{amsmath}
\usepackage{comment}
\usepackage{caption,subcaption}     
\usepackage[figuresright]{rotating} 

\usepackage{algorithmic}            
\usepackage{algorithm}              

\usepackage{multirow}               
\usepackage[table,xcdraw]{xcolor}   
\usepackage{arydshln}               

\usepackage{amsthm}                 
\newtheorem{theorem}{Theorem}[]          
\newtheorem{Proposition}[theorem]{Proposition}  

\journal{ArXiv}

\begin{document}

\begin{frontmatter}

\title{Mixed-model Sequencing with Stochastic Failures: A Case Study for Automobile Industry}
\tnotetext[t1]{This research did not receive any specific grant from funding agencies in the public, commercial, or
not-for-profit sectors.}

\author[1]{I. Ozan Yilmazlar\corref{cor1}}
\ead{iyilmaz@clemson.edu}

\author[1]{Mary E. Kurz}
\ead{mkurz@clemson.edu}

\author[1]{Hamed Rahimian}
\ead{hrahimi@clemson.edu}

\cortext[cor1]{Corresponding author}
\address[1]{Department of Industrial Engineering, Clemson University, Clemson SC, 29634}

\begin{abstract}
    In the automotive industry, the sequence of vehicles to be produced is determined ahead of the production day. However, there are some vehicles, failed vehicles, that cannot be produced due to some reasons such as material shortage or paint failure. These vehicles are pulled out of the sequence, and the vehicles in the succeeding positions are moved forward, potentially resulting in challenges for logistics or other scheduling concerns. 
    This paper proposes a two-stage stochastic program for the mixed-model sequencing (MMS) problem with stochastic product failures, and provide improvements to the second-stage problem. To tackle the exponential number of scenarios, we employ the sample average approximation approach and two solution methodologies. On one hand, we develop an L-shaped decomposition-based algorithm, where the computational experiments show its superiority over solving the deterministic equivalent formulation with an off-the-shelf solver. Moreover, we provide a tabu search algorithm in addition to a greedy heuristic to tackle case study instances inspired by our car manufacturer partner. Numerical experiments show that the proposed solution methodologies generate high quality solutions by utilizing a sample of scenarios. Particularly, a robust sequence that is generated by considering car failures can decrease the expected work overload by more than 20\% for both small- and large-sized instances.
\end{abstract}

\begin{keyword}
Stochastic programming \sep Mixed-model sequencing \sep Branch-and-Benders-cut \sep Heuristics \sep Tabu Search
\end{keyword}
\end{frontmatter}

\section{Introduction}\label{section: mms-introduction}
Mixed-model assembly lines (MMAL) are capable of producing several configurations, models, of a product. The number of models increases drastically as the complexity and customizability of the product expand. The number of theoretical configurations of vehicles from a German car manufacturer is up to $10^{24}$ \citep{pil2004linking}. Different configurations require distinct tasks at each station which induces high variation in the processing times, though each station has a fixed maximum time available. In fact, station workload is distributed through line balancing such that each station's \emph{average} workload conforms to this maximum time. When a station has more work allocated to it for a particular model (\emph{work overload}), interventions are needed to maintain the flow of products in the assembly line, thereby avoiding line stoppages. Interventions can be considered in advance, through sequencing decisions, or at the time of disruption, through utility workers. When these interventions fail, the line may stop production until the situation is resolved. Thus, it is essential to distribute the high work-load models along the planning horizon to avoid line stoppages. 

The mixed-model sequencing (MMS) problem sequences products in an MMAL to minimize work overload at the stations. Data from our car manufacturing partner shows that the variation in processing times is high when customization appears on a main part, e.g., engine type: electric, diesel, gasoline, or hybrid. Car manufacturers have adapted their assembly lines for the mixed-model production of vehicles with diesel and gasoline engines. However, the assembly of electric vehicles (EV) in the same line has brought new challenges, while not eliminating the production of vehicles with diesel or gasoline engines. Unlike other vehicles, electric and hybrid vehicles have large batteries which causes a huge difference in tasks, e.g., at the station where the battery is loaded. As the proportion of electric and hybrid vehicles grows in a manufacturer's mix, the impact of supply problems increases. Sometimes, a part is delayed from a supplier, so a designed  sequence of vehicles will have a missing vehicle. Even if this vehicle has a gasoline or diesel engine, its absence may impact the battery-intensive stations. As a manufacturer's mix of vehicles grows more specialized with more time-consuming content for a large subset, without alternative tasks for the vehicles without the specialized content, the impact of missing vehicles on a  carefully designed sequence grows.

\begin{figure}[h]
\centering
\begin{subfigure}{0.5\textwidth}
  \centering
  \includegraphics[width=0.85\linewidth]{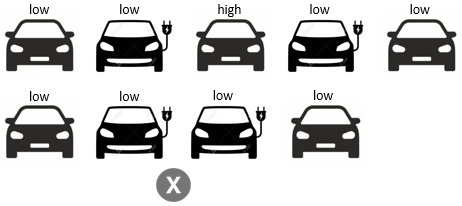}
  \caption{non-robust sequence}
  \label{fig_sub: non-robust sequenc}
\end{subfigure}%
\begin{subfigure}{0.5\textwidth}
  \centering
  \includegraphics[width=0.85\linewidth]{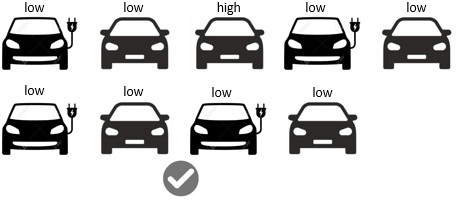}
  \caption{robust sequence}
  \label{fig_sub: robust sequence}
\end{subfigure}
\caption{Illustration of a non-robust and robust sequence to stochastic failures}
\label{fig: robust sequence}
\end{figure}

Some vehicles in a production sequence may not be ready for assembly on the production day for various reasons, such as the body not being ready, paint quality issues, or material shortage. Such vehicles, referred to as \emph{failed vehicles}, need to be pulled out of the sequence. The resulting gap is closed by moving the succeeding vehicles forward. This process and the resulting additional work overload occurrence is illustrated in Figure \ref{fig: robust sequence} for a battery loading station. The processing time at this station is longer than the cycle time for EVs and shorter than the cycle time for non-EVs, and assume that back-to-back EVs cause work overload. We schedule five vehicles, two electric and three non-electric. One of the non-EVs (third in both scheduled  sequences) has a high failure probability. The initial sequences with no failures, while different, both will lead to no work overload. Assuming the third vehicle fails, we have different consequences for the resultant sequence of vehicles. In the non-robust sequence, removing the failed non-EV results in two EVs in a row, which will cause a work overload. However, the robust sequence, which is composed of the same vehicles in a different order, can withstand the failure of the third vehicle without causing a work overload. We refer to this sequence as the  ``robust'' sequence because no work overload occurs when the vehicle with high failure probability is pulled out of the sequence.

In this study, we generate robust sequences that consider the vehicles' potential failures to reduce additional work overloads. We focus on the final assembly line, assuming that vehicles follow the same sequence as they arrive from the paint shop and resequencing is not an option; when a vehicle is removed from the sequence, the following vehicles close the gap. The contributions of this study are as follows:

\begin{itemize}
\item We provide a two-stage stochastic program for a MMS problem with stochastic product failures, and we provide improvements to the second-stage problem. To the best of our knowledge, this is the first study that considers stochastic failures of products in MMS.
\item We adopt the sample average approximation (SAA) approach to tackle the exponential number of scenarios. The numerical experiments show that we can generate robust solutions with an optimality gap of less than 1\% and 5\% by utilizing a sample of scenarios, for the small-sized and industry-sized instances, respectively.
\item We develop an L-shaped decomposition-based algorithm to solve small-sized instances. The numerical results show that the L-shaped algorithm outperforms an off-the-shelf solver, solving the deterministic equivalent formulation (DEF), in terms of both quality and computational time. 
\item To solve industry-sized instances, we propose a greedy heuristic and a tabu search (TS) algorithm that is accelerated in convergence with problem-specific tabu rules, and in objective reevaluation each time a new solution is visited.
\item We conduct a case study with the data inspired by our car manufacturer industrial partner. The numerical experiments show that we can reduce the work overload by more than 20\% by considering stochastic car failures and solving the corresponding problem with the proposed solution methodologies.
\end{itemize}

The remainder of this paper is structured as follows. MMS related literature is reviewed in Section \ref{section: mms-related-work}. The tackled problem is defined, and the mathematical formulation of the proposed problem is presented in Section \ref{section: mms-problem-definition}. Exact and heuristic solution approaches in addition to the SAA approach are presented in Section \ref{section:mms-solution-approaches}. In Section \ref{section: mms-computational-experiments}, we execute numerical experiments to analyze the performance of proposed solution methodologies and present the results. Finally, a summary of our findings and a discussion about future work are given in Section \ref{section: mms-conclusion}.

\section{Related Work}\label{section: mms-related-work}

Manufacturers use various design configurations of MMALs to maximize their revenue. The optimization process of the assembly line sequencing takes these design configurations into account. The first paper that articulates the MMS was presented by \citet{kilbridge1963assembly}. The researchers tackle the MMS with varied characteristics which required a systematic categorization of the components and the operating system of MMS problems. \citet{dar1978mixed} categorizes the MMAL into four categories based on their main characteristics of assembly lines: product transfer system, product mobility on the conveyor, accessibility among adjacent stations, and the attribute of the launching period. An analytic framework for the categorization of \citet{dar1978mixed} is given by \citet{bard1992analytic}. Later, a survey is presented by \citet{boysen2009sequencing}, where they define tuple notation for the sequencing problems based on more detailed characteristics of assembly lines, including work overload management, processing time, concurrent work, line layout, and objective in addition to the main characteristics.

Several objectives are employed to evaluate the performance of the assembly line sequence. The most common objective in the literature, also adopted in this study, is minimizing the total work overload duration, proposed by \citet{yano1991sequencing}. \citet{tsai1995mixed} describes hiring utility workers to execute tasks so that production delays are avoided, which leads to the objective of minimizing the total utility work duration. \citet{fattahi2009sequencing} minimize the total idle time in addition to utility work. \citet{boysen2011sequencing} propose minimizing the number of utility workers instead of the total utility work duration in order to improve utility worker management. 

A few exact solution methods are proposed in the literature to solve the deterministic MMS problem. \citet{scholl1998pattern} proposes a decomposition approach that uses patterns of different sequences, called pattern-based vocabulary building. They use a column generation method to solve the linear relaxation of the formulation and an informed tabu search is adapted to determine the pattern sequence. \citet{bolat2003mathematical} propose a job selection problem that is solved prior to the sequencing problem. They employ a due date-oriented cost function as an objective, and the work overload is restricted as a hard constraint. They develop a branch-and-bound (B\&B) algorithm that is improved with some dominance criteria, a procedure to compare sequences based on the quality, which can select 50 jobs out of 100 in seconds. \citet{kim2007product} present a B\&B algorithm to solve the MMS problem with sequence-dependent setup times. They calculate a lower bound on the work overload of the current sequence and the minimum possible work overload of the unconsidered configurations. The model can solve instances with up to five stations and 30 configurations. \citet{boysen2011sequencing} integrate a skip policy for utility work into the MMS and formulate the new problem as an mixed-integer linear program (MILP). They propose a B\&B algorithm with improved lower bound calculations and dominance rules.

There are several heuristic and meta-heuristic approaches related to the MMS. The most popular algorithm is the genetic algorithm (GA) which is adapted in several ways to solve MMS problems \citep{hyun1998genetic, kim1996sequencing, ponnambalam2003genetic, leu1996sequencing, celano1999evolutionary, akgunduz2010adaptive, zhang2020multi}. \citet{akgunduz2011review} review GA-based MMS solution approaches. Other popular evolutionary algorithms that are used to solve MMS include ant colony optimization \citep{akpinar2013hybridizing, zhu2011ant, kucukkoc2014simultaneous}, particle swarm \citep{rahimi2007new, wei2011hybrid}, scatter search algorithm \citep{rahimi2007multi, cano2010scatter, liu2014advanced}, and imperialist competitive algorithm  \citep{zandieh2019imperialist}. 


While the majority of the MMS literature focuses on models with deterministic parameters, there are a few studies that consider stochastic parameters on either processing times or demand. The seminal study with stochastic processing times is proposed by \citet{zhao2007modeling}. They provide a Markov chain based approach that minimizes the expected work overload duration. This approximation is done by generating sub-intervals of possible positions of workers within the stations. The expected work overload is calculated based on the lower and upper bounds of the intervals. \citet{mosadegh2017heuristic} propose a heuristic approach, inspired by Dijkstra's algorithm, to tackle a single-station MMS with stochastic processing times. They formulate the problem  as a shortest path problem. \citet{mosadegh2020stochastic} formulate a multiple-station MMS with stochastic processing times as an MILP. They provide a Q-learning-based simulated annealing (SA) heuristic to solve industry-sized problems and show that the expected work overload is decreased compared to the deterministic problem. \citet{brammer2022stochastic} propose a reinforcement learning approach to solve MMS by negatively rewarding work overload occurrences. They show that the proposed approach provides at least 7\% better solutions than SA and GA. Moreover, stochastic parameters are considered in integrated mixed-model balancing and sequencing problems as well: in processing times \citep{agrawal2008collaborative, ozcan2011genetic, dong2014balancing} and in demand \citep{sikora2021benders}. 

Although numerous studies have been conducted on the sequencing problems, only \citet{hottenrott2021robust} consider the product failures in sequence planning, yet in the car sequencing structure. To the best of our knowledge, there is no research available that establishes robust sequences in the MMS structure that can withstand work overloads caused by product failures.

\section{Problem Statement and Mathematical Formulation}\label{section: mms-problem-definition}
In Section \ref{section: mms-problem-statement}, we define the MMS with stochastic failures and illustrate the problem with an example. Then, in Section \ref{section: mms-mathematical-model}, we provide a two-stage stochastic program for our problem.

\subsection{Problem Statement}\label{section: mms-problem-statement}
In an MMAL, a set of workstations are connected by a conveyor belt. Products, launched with a fixed rate, move along the belt at a constant speed of one time unit (TU). The duration between two consecutive launches is called the cycle time $c$, and we define the station length $l_k\ge c$ in TU as the total TU that the workpiece requires to cross the station $k \in K$. Operators work on the assigned tasks and must finish their job within the station length, otherwise, the line is stopped or a so-called utility worker takes over the remaining job. The excess work is called {\it work overload}. The sequence of products therefore has a great impact on the efficiency of the assembly line. MMS determines the sequence of a given set of products $V$ by assigning each product $v \in V$ to one of the positions $t \in T$. 

Formulating the MMS problem based on the vehicle configurations instead of vehicles is usual \citep{bolat1992scheduling, bard1992analytic, scholl1998pattern}, however, automobile manufacturers offer billions of combinations of options \citep{pil2004linking}. When this high level of customization is combined with a short lead time promised to the customers, each vehicle produced in a planning horizon becomes unique. In this study, the vehicles are sequenced instead of configurations since the case study focuses on an automobile manufacturing facility with a high level of customization. In order to do so, we define a binary decision variable $x_{vt}$, which takes value of  1 if vehicle $v \in V$ is assigned to position $t \in T$. The processing time of vehicle $v \in V$ at station $k \in K$ is notated by $p_{kv}$. The starting position and work overload of the vehicle at position $t \in T$ for station $k \in K$ are represented by $z_{kt}$ and $w_{kt}$, respectively. Table \ref{tbl: notations} lists all the parameters and decision variables used in the proposed model. While second-stage decision variables are scenario-dependent, we drop such dependency for notation simplicity throughout the paper unless it is needed explicitly for clarity. 

\begin{table}[]
\centering
\scalebox{0.75}{
\begin{tabular}{
>{\columncolor[HTML]{FFFFFF}}l 
>{\columncolor[HTML]{FFFFFF}}l lll}
\textbf{Sets and Index}     & { \textbf{}}                                                                               &  &  &  \\ \cline{1-2}
$V, v$                          & Vehicles                                                                                      &  &  &  \\
$K, k$                          & Stations                                                                                      &  &  &  \\
$T, t$                          & Positions                                                                                        &  &  &  \\
$\Omega, \omega$                          & Scenarios                                                                                     &  &  &  \\
                              &                                                                                               &  &  &  \\
\textbf{Parameters}           & \textbf{}                                                                                     &  &  &  \\ \cline{1-2}
$p_{kv}$                           & The processing time of vehicle $v \in V$ at station $k \in K$                                           &  &  &  \\
$l_k$                            & The length of station $k \in K$                                                                        &  &  &  \\
$c$                             & The cycle time                                                                                &  &  &  \\
$f_v$                            & The failure probability of vehicle $v \in V$                                                          &  &  &  \\
$e_{v\omega}$                           & 1 if vehicle $v \in V$ exists at scenario $\omega \in \Omega$, 0 otherwise                                          &  &  &  \\
                              &                                                                                               &  &  &  \\
\textbf{First-Stage Decision   variables} & \textbf{}                                                                                     &  &  &  \\ \cline{1-2}
$x_{vt}$                           & 1 if vehicle $v \in V$ is assigned to position $t \in T$, 0 otherwise                                        &  &  &  \\
                      &  &  &  \\
\textbf{Second-Stage Decision   variables} & \textbf{}                                                                                     &  &  &  \\ \cline{1-2}
                      
$w_{kt}$                           & The work overload at station $k \in K$ at position $t \in T$                                     &  &  &  \\
$z_{kt}$                           & Starting position of operator at station $k \in K$ at the beginning of position $t \in T$  &  &  &  \\
$b_{kt}$                           & The processing time at station $k \in K$ at position $t \in T$                       &  &  & 
\end{tabular}}
\caption{List of parameters and decision variables used in the model}
\label{tbl: notations}
\end{table}

In this paper, we adopt the side-by-side policy as a work overload handling procedure. A utility worker is assumed to work with the regular worker side-by-side, enabling work to be completed within the station borders. The objective of MMS with the side-by-side policy is to minimize the total duration of work overloads, i.e., the total duration of the remaining tasks that cannot be completed within the station borders. The regular operator stops working on the piece at the station border so they can start working on the next workpiece at position $l_k - c$ in the same station.

\begin{figure}[h]\vspace*{4pt}
\captionsetup{justification=centering}
    \centerline{\includegraphics[scale=0.5]{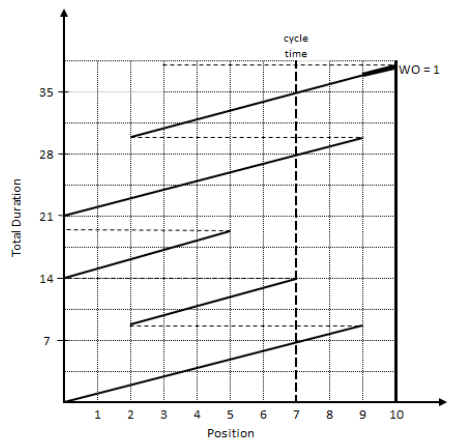}}
    \caption{Illustration of mixed-model assembly line with five vehicles, cycle time $c=7$, and station length $l_k=10$. From bottom to top, the diagonal lines correspond to vehicle configurations A, B, B, A, A}
    \label{fgr:mms}
\end{figure}

We illustrate an MMAL with the side-by-side policy in Figure \ref{fgr:mms} which represents a station that processes five vehicles. The left and right vertical bold lines represent the left and right borders of the station. Assume that the cycle time $c$ is 7 and the station length $l_k$ is 10 TU, i.e., it takes 10 TU for the conveyor to flow through the station.  This specific station processes two different configurations of vehicles: configurations A and B. While configurations A requires option 1, configurations B does not, so the processing times of configurations A and  B are 9 and 5, respectively. Figure \ref{fgr:mms} illustrates the first five vehicles in the sequence which is [A, B, B, A, A]. The diagonal straight lines represent the position of the vehicle in the station. The worker always starts working on the first vehicle at position zero, left border of the station. The second vehicle is already at position $2=9-c$ when the worker completes working on the first vehicle. Note that the next vehicle enters the station borders a cycle time after the current vehicle enters the station borders. The tasks of the second vehicle are completed when the third vehicle has just entered the station. The worker has 2 TU of idle time when the tasks of the third vehicle are completed. The worker starts working on the fifth vehicle on position 2 and the processing time of the fifth vehicle is 9 TU which causes a work overload of 1 TU, $2+9-l_k=1$. The job of processing these five vehicles could not be completed within the station borders but with the help of a utility worker, we assume that the job is completed at the station border at the cost of 1 TU work overload. The worker will continue working on the sixth vehicle at position $3=l_k-c$, and this process continues.

We note that the vehicles usually go  through the body shop and paint shop in the scheduled sequence before the assembly process. Hence, the failed vehicles must be pulled out of the sequence, and its position cannot be compensated, i.e.,  resequencing is not an option. It is assumed that each vehicle $v \in V$ (with its unique mix of configurations) has a failure probability $f_v$, and failures are independent of each other. The failures are related to the specifications of the vehicle, e.g., the increased production rate of EVs may induce higher failure rates, or painting a vehicle to a specific color may be more problematic. In our numerical experiments in Section \ref{section: mms-computational-experiments}, we estimate the failure probabilities from the historical data by doing feature analysis and using logistic regression. 

\subsection{Mathematical Model Under Uncertainty}\label{section: mms-mathematical-model}
In this section, first, we provide a two-stage stochastic program for our problem. Next, we discuss improvements of the proposed formulation.

Motivated by the dynamics of an MMAL, we formulate our problem as a two-stage stochastic program. The sequence of vehicles is decided in the first stage (here-and-now), before the car failures are realized. Once the car  failures are realized, the work overload is minimized by determining the second-stage decisions (wait-and-see) given the sequence. First-stage decisions are determined by assigning each vehicle to a position such that the expected work overload in the second stage is minimized. 

To formulate the problem, suppose that various realizations of the car failures are  represented by a collection of finite scenarios $\Omega$. As each vehicle either exists or fails at a scenario, we have a total of $2^{|V|}$ scenarios. We let $\Omega=\{\omega_1,\ldots,\omega_{2^{|V|}}\}$, with $\omega$ indicating a generic scenario. To denote a scenario $\omega$, let $e_{v\omega}=1$ if vehicle $v$ exists and $e_{v\omega}=0$ if  vehicle $v$ fails at  scenario $\omega \in \Omega$. We can then calculate the probability of scenario $\omega$ as $\rho_\omega=\prod_{v=1}^{|V|}f_v^{1-e_{v\omega}} (1-f_v)^{e_{v \omega}}$ such that $\sum_{ \omega \in \Omega}\rho_\omega=1$, where $f_v$ denotes the failure probability of vehicle $v \in V$. 

A two-stage stochastic program for the \textit{full-information problem},  where all possible realizations are considered, is as follows:
\begin{subequations}
\label{form: 1st}
\begin{flalign}
\min_{x} \quad & \sum_{\omega\in \Omega}\rho_\omega Q(x,\omega) \label{c1.0} \\
\text{s.t.} \quad & \sum_{v\in V} x_{vt} = 1, \qquad    t\in T \label{c1.1} \\ 
& \sum_{t\in T} x_{vt} = 1, \qquad    v\in V \label{c1.2} \\ 
& x_{vt} \in \{0,1\}, \qquad    t\in T, \enspace   v\in V \label{c1.3}
\end{flalign}
\end{subequations}
where 
\begin{subequations}
\label{form: 2nd}
\begin{flalign}
Q(x,\omega) = \min_{z,w,b} \quad & \sum_{k\in K}\sum_{t\in T_{\omega}}w_{kt}  \label{c2.0} \\
\text{s.t.} \quad & b_{kt} = \sum_{v\in V} p_{kv}x_{vt}, \qquad   k\in K, \enspace   t\in T_{\omega} \label{c2.1} \\ 
& z_{kt} - z_{k(t+1)} - w_{kt}  \leq c - b_{kt}, \qquad   k\in K, \enspace   t\in T_{\omega}, \enspace \label{c2.2} \\ 
& z_{kt} - w_{kt} \leq l_k - b_{kt}, \qquad   k\in K, \enspace   t\in T_{\omega}, & \label{c2.3} \\
& z_{k0} = 0, \qquad    k\in K, \enspace \label{c2.4} \\ 
& z_{k(|T_{\omega}|+1)} = 0, \qquad    k\in K, \enspace  \label{c2.5} \\ 
& z_{kt}, w_{kt} \geq 0, \qquad   k\in K, \enspace   t\in T_{\omega}, \enspace \label{c2.6} 
\end{flalign}
\end{subequations}
In the first-stage problem \eqref{form: 1st}, the objective function represents the expected work overload, i.e., the cost associated with the second-stage problem. Constraint sets \eqref{c1.1} and \eqref{c1.2} ensure that exactly one vehicle is assigned to each position and each position has exactly one vehicle. respectively. Constraint set \eqref{c1.3} presents the domain of the binary first-stage variable.
The second-stage problem \eqref{form: 2nd} minimizes the total work overload throughout the planning horizon, given the sequence and scenario $\omega \in \Omega$. Note that 
 $T_{\omega}$ denotes the set of positions of non-failed vehicles at scenario $\omega \in \Omega$, which is obtained by removing failed vehicles. Constraint set \eqref{c2.1} determines the processing time $b_{kt}$ at station $k$ at position $t$. The starting position and workload of the vehicles at each station are determined by constraint sets \eqref{c2.2} and \eqref{c2.3}, respectively. The constraint set \eqref{c2.4} ensures that the first position starts at the left border of the station. Constraint set \eqref{c2.5} builds regenerative production planning, in other words, the first position of the next planning horizon can start at the left border of the  station. The constraint set \eqref{c2.6} defines the second-stage variables as continuous and nonnegative.

Several remarks are in order regarding the proposed two-stage stochastic program \eqref{form: 1st} and \eqref{form: 2nd}. 
First, the number of decision variables and the set of constraints in \eqref{form: 2nd} are scenario-dependent, as the valid positions $T_{\omega}$ are obtained based on the failure scenario $\omega \in \Omega$.  Second, the proposed two-stage stochastic program \eqref{form: 1st} and \eqref{form: 2nd} has a simple recourse. That is, once the sequence is determined and the failures are realized, the work overloads are calculated from the sequence of the existing vehicles, without resequencing. 

In the remainder of this section, we first provide two modified models for the second-stage problem so that the number of decision variables and the set of constraints are no longer scenario-dependent. Then, we provide two monolithic MILP formulations for the deterministic equivalent formulation (DEF) of the two-stage stochastic program of MMS with stochastic failures.

\begin{figure}[h]
\centering
\begin{subfigure}{0.33\textwidth}
  \centering
    \includegraphics[width=1.0\textwidth]{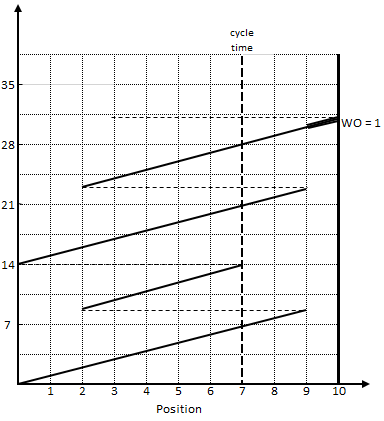}\hfill
      \caption{Original model}
  \label{fig_sub: non_linear_model}
\end{subfigure}%
\begin{subfigure}{0.33\textwidth}
  \centering
    \includegraphics[width=1.0\textwidth]{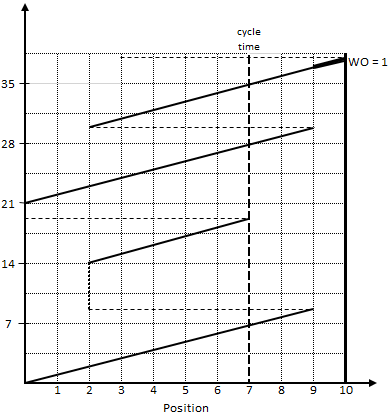}\hfill
  \caption{(Modified) standard  model}
  \label{fig_sub: standard_model}
\end{subfigure}%
\begin{subfigure}{0.33\textwidth}
  \centering
    \includegraphics[width=1.0\textwidth]{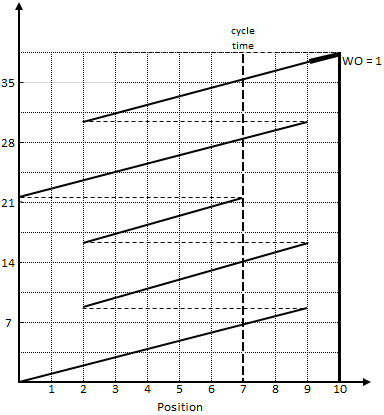}
  \caption{(Modified) improved  model}
  \label{fig_sub: improved_model}
\end{subfigure}%
\caption{Assembly line illustration of proposed models}
\label{fig: models_illustration}
\end{figure}

For each scenario, we modify the second-stage problem by updating the processing times of failed vehicles  instead of removing the failed vehicles. In Figure \ref{fig: models_illustration}, we demonstrate how the original model \eqref{form: 2nd} and modified models represent car failures. To this end, we consider the example given in Figure \ref{fgr:mms} and assume that the vehicle at the second position fails. In the original model, the failed vehicles are removed from the sequence and the succeeding vehicles moved forward (Figure \ref{fig_sub: non_linear_model}). The proposed modified models, referred to as {\it standard model} and {\it improved model}, are explained below.
In Section \ref{section: mms-exact-results}, we discuss the impact of these modified models on the computational time and solution quality.  

\textbf{Standard Model:} In a  preprocessing step, the processing time of vehicle $v$ is set to zero for all stations if the vehicle fails in scenario $\omega$. Accordingly, since this modification is conducted by adding  uncertainty to the  processing times, the scenario index $\omega$ is added to the processing time. That is, $e_{v\omega}=0 \Rightarrow p_{kv\omega}=0 \enspace   k\in K$. 
Based on this modification, the second-stage problem for a given scenario $\omega$ can be presented as 
\begin{subequations}
\label{form: 2nd_standard}
\begin{flalign}
Q(x,\omega)= \min_{w,z,b} \quad & \sum_{k\in K}\sum_{t\in T}w_{kt}  \label{c3.0} \\
\text{s.t.} \quad & b_{kt} = \sum_{v\in V} p_{kv}x_{vt}, \qquad   k\in K, \enspace   t\in T, \enspace  & \label{c3.3} \\ 
& z_{kt}+b_{kt}-w_{kt}-z_{k(t+1)} \leq c, \qquad   k\in K, \enspace   t\in T, \enspace & \label{c3.4} \\ 
& z_{kt}+b_{kt}-w_{kt} \leq l_k, \qquad   k\in K, \enspace   t\in T, \enspace & \label{c3.5} \\ 
& z_{k0} = 0, \qquad   k\in K, \enspace  & \label{c3.6} \\ 
& z_{k(T+1)} = 0, \qquad   k\in K, \enspace  & \label{c3.7} \\ 
& z_{kt}- \beta_k b_{kt}-z_{k(t+1)} \leq 0, \qquad   k\in K, \enspace   t=\{1,\ldots, T-1\} & \label{c3.8} \\ 
& z_{kT} - w_{kT} - \beta_k b_{kT} \leq 0, \qquad   k\in K, \enspace  & \label{c3.9} \\ 
& z_{kt}, w_{kt}, b_{kt} \geq 0, \qquad   k\in K, \enspace   t\in T, \enspace  & \label{c3.11} 
\end{flalign}
\end{subequations}
The objective function \eqref{c3.0} and the constraints \eqref{c3.3}--\eqref{c3.7} are the same as in formulation \eqref{form: 2nd}, except the set of positions and the length of the sequence are not scenario-dependent anymore. Constraint set \eqref{c3.8} guarantees that the starting position at station $k$ at position $t+1$ equals the starting position of position $t$ when the vehicle assigned to position $t$ fails. Constraint set \eqref{c3.9} assures that the regenerative production planning is kept in case the vehicle at the end of the sequence fails. The parameter $\beta$ is calculated in a way that $\beta_k b_{ktn}>z_{ktn}$ which makes constraint sets \eqref{c3.8} and \eqref{c3.9} non-effective for the positions that have existing vehicles. Hence, $\beta_k$ equals the maximum possible starting position divided by the minimum processing time for station $k$,  $\beta_k = \frac{l_k-c}{\min_{v \in V}\{p_{kv}\}}>0$. Note that the processing times in this calculation are the actual processing times before the preprocessing step. Also, $\beta_k $ is well-defined as the the minimum processing time is strictly greater than zero.
Figure \ref{fig_sub: standard_model} demonstrates that in the standard model, the processing time of the second vehicle is set to zero, so the operator starts working on the third vehicle at position two where the operator was going to start working on the second vehicle if it had not failed. 

\textbf{Improved Model:} In order to reduce the size of the standard model, we modify this model as follows. During the preprocessing step, the processing time of vehicle $v$ is set to the cycle time for all stations if a vehicle fails at scenario $\omega$. Let us refer to  the vehicles with processing time equal to cycle time for all stations as ``neutral" because these vehicles do not have any impact on the schedule in terms of work overload (see Proposition \ref{pp: neutral vehicles} and its proof). 
In other words, we transform failed vehicles into \textit{neutral} vehicles, i.e., $e_{v\omega}=0 \Rightarrow p_{kv\omega}=c \enspace k\in K$.

\begin{Proposition}\label{pp: neutral vehicles}
A neutral vehicle has the same starting position as its succeeding vehicle at all stations. That is, $ b_{kt}=c \enspace \Rightarrow \enspace z_{k(t+1)} = z_{kt}$. 
\begin{proof}
    The operator's starting position of the vehicle at $t+1$ is $z_{k(t+1)} = z_{kt}+b_{kt}-c-w_{kt}$. Assume that the vehicle at position $t$ is a neutral vehicle. We have $z_{k(t+1)} = z_{kt}-w_{kt}$. Hence, showing that the neutral vehicles never cause a work overload, $w_{kt}=0,$  completes the proof. We know that the maximum starting position at a station is $\max_{t \in T}\{z_{kt}\}=l_k-c$, which is a result of two extreme cases: an operator finishes working on a workpiece at the right border of a  station or the operator cannot finish the work so we have a work overload. The starting position is less than $l_k-c$ for other cases. Therefore, a vehicle with a processing time less than or equal to $c$ at a station cannot cause any work overload. This completes the proof.
\end{proof}
\end{Proposition}

As a result of Proposition \ref{pp: neutral vehicles},  constraints \eqref{c3.8} and \eqref{c3.9} can be removed from the standard model. Hence, the problem size is reduced. 
Figure \ref{fig_sub: improved_model} contains an illustration for Proposition \ref{pp: neutral vehicles}. The second vehicle becomes neutral when its processing time is set to cycle time so that the third vehicle starts at the same position as the second vehicle. 

Using the standard or improved model, the DEF for MMS with stochastic failures can be obtained by adding the first-stage constraints \eqref{c1.1}--\eqref{c1.3} to the  corresponding second-stage formulation, and by adding copies of  all second-stage variables and constraints. We skip the details for brevity.

\section{Solution Approaches} \label{section:mms-solution-approaches}
In Sections \ref{section: mms-exact-solution} and \ref{section: mms-heuristics}, we  propose an L-shaped decomposition-based algorithm and a tabu search algorithm in addition to a greedy heuristic, respectively, to solve the models presented in Section \ref{section: mms-mathematical-model}. Then, in Section \ref{mms: solution quality assessment}, the SAA approach is motivated and a solution quality assessment scheme is presented. 

\subsection{Exact Solution Approach }\label{section: mms-exact-solution}
 For the ease of exposition, we consider an abstract formulation of the two-stage stochastic program presented in Section \ref{section: mms-mathematical-model} as follows:
\begin{flalign}\label{eq:true stochastic problem}
& z^*= \min_{x \in X}  \mathbb{E}[Q(x,\xi_\omega)], 
\end{flalign}
where $x$ denotes the first-stage decisions variables and $X:=\{x \in \{0,1\}^{|V|\times |T|} \colon Ax=b\}$ is the feasible region of decision variables $x$, i.e., the set of points satisfying constraints \eqref{c1.1} - \eqref{c1.3}. Moreover, we represent the second-stage problem for the standard or improved model, presented in Section \ref{section: mms-mathematical-model}, as 
\begin{equation}
\label{eq: 2nd_gen}
    Q(x,\xi_\omega) = \min_{y}\{q^{\top}y | Dy \ge h_\omega - T_\omega x, \; y \ge 0\}, 
\end{equation}
where $y$ represents the second-stage decision variables and $\xi_\omega=(h_\omega, T_\omega)$. The expectation of the recourse problem becomes $\mathbb{E}[Q(x,\xi_\omega)] = \sum_{\omega\in \Omega}\rho_\omega Q(x,\xi_\omega)$. 

The L-shaped method is a procedure that has been successfully used to solve large-scale two-stage stochastic programs.  
Note that for any $\omega \in \Omega$, function $Q(x,\xi_\omega)$, defined in \eqref{eq: 2nd_gen}, is convex in $x$ because $x$ appears on the right-hand side of constraints. Hence, we propose to  iteratively construct its underestimator. To this end, for each $\omega \in\Omega$ and a given first-stage decision $x \in X$, we consider a  {\it subproblem} that takes the form of \eqref{eq: 2nd_gen}. Moreover, we create a {\it relaxed master problem}, which contains a partial, but increasingly improving, representation of $Q(x,\xi_\omega)$,  for each $\omega \in \Omega$,  through the so-called \textit{Benders' cuts}. 
Recall that our proposed two-stage stochastic programs have a relative complete recourse, that is, for any first-stage decision $x$, there is a feasible second-stage variable $y$. Thus, an underestimator of $Q(x,\xi_\omega)$ can be constructed by only the so-called {\it Benders' optimality cuts}.  

We now describe more details on our proposed L-shaped  algorithm. We form the relaxed master problem for formulation  \eqref{eq:true stochastic problem} and \eqref{eq: 2nd_gen} as follows: 
\begin{subequations}
\label{form: MP}
\begin{flalign} 
\min_{x,\theta} \quad & \sum_{\omega \in \Omega} \rho_\omega \theta_{\omega}  \label{MP:c0} \\
\text{s.t.} \quad & x \in X \label{MP:c1}  \\
& \theta_{\omega} \geq   G^{\iota}_\omega x  + g^{\iota}_\omega, \quad    \iota \in \{1,\ldots,l\},  \enspace   \omega \in \Omega, \label{MP:c2} 
\end{flalign}
\end{subequations}
where the auxiliary variable $\theta_\omega$ approximates the optimal value of the second-stage problem under scenario $\omega \in \Omega$, i.e., $Q(x, \xi_\omega)$, through cuts $\theta_{\omega} \geq   G^{\iota}_\omega x  + g^{\iota}_\omega$ formed up to iteration $l$. 

Let $(\hat{x}^{\iota}, \hat{\theta}^{\iota})$ be an optimal solution to the relaxed master problem \eqref{form: MP}. For each scenario $\omega \in \Omega$, we form a subproblem \eqref{eq: 2nd_gen} at $\hat{x}^{\iota}$. Suppose that given  $\hat{x}^{\iota}$, $\hat{\pi}_{\omega}^{\iota}$ denotes an optimal dual vector associated with the constraints in \eqref{eq: 2nd_gen}. That is, $\hat{\pi}_{\omega}^{\iota}$ is an optimal extreme point of the dual subproblem (DSP) 
\begin{equation}
\label{eq: 2nd_gen_dual}
    \max_{\pi}\{\pi_{\omega}^{\top}(h_\omega - T_\omega \hat{x}^{\iota}) | \pi_{\omega}^{\top} D \le q^{\top}, \; \pi_{\omega} \ge 0\}, 
\end{equation}
where $\pi_{\omega}$ is the associated dual vector. 
Then, using linear programming duality, we generate an optimality cut as
\begin{equation}
\label{eq:opt-stoc}
    \theta_\omega \ge G^{\iota}_\omega x  + g^{\iota}_\omega,
\end{equation}
where $G^{\iota}_\omega= -  (\hat{\pi}_{\omega}^{\iota})^{\top} T_\omega$ and $g^{\iota}_\omega=  (\hat{\pi}_{\omega}^{\iota} )^{\top} h_\omega$. 

Our proposed L-shaped algorithm iterates between solving the relaxed master problem \eqref{form: MP} and  subproblems \eqref{eq: 2nd_gen}  (one for each $\omega \in \Omega$) until a convergence criterion on the upper and lower bounds  is satisfied. 
This algorithm results in an L-shaped method with {\it multiple} cuts. 

In order to exploit the specific structure of the MMS problem and to provide improvements on the dual problem, let us define variables $\pi^{sp}$, $\pi^{wo}$, $\pi^{fs}$, $\pi^{ch}$, $\pi^{sf}$, and $\pi^{cf}$ corresponding to starting position constraints \eqref{c3.4}, work overload constraints \eqref{c3.5}, first station starting position constraints \eqref{c3.6}, regenerative production planning constraints \eqref{c3.7}, starting position of the vehicles following a failed vehicle \eqref{c3.7}, and regenerative production planning with failed vehicles \eqref{c3.8}, respectively. The DSP for scenario $\omega \in \Omega$ at a candidate solution $\hat{x}^{\iota}$, obtained by solving a relaxed master problem,  can be formulated as follows:
\begin{subequations}
\label{form: dual-standard}
\begin{flalign}
\max_{\pi} \quad & \sum_{k\in K}\sum_{t\in T} \pi^{sp}_{kt} (\sum_{v\in V} p_{kv}\hat{x}_{vt} - c) + \pi^{wo}_{kt} (\sum_{v\in V} p_{kv}\hat{x}_{vt} - l_{k}) & \label{c5.0} \\
\text{s.t.} \quad & \pi^{sp}_{k0} + \pi^{wo}_{k0} + \pi^{fs}_{k} + \pi^{sf}_{k0} \leq 0, \qquad   k\in K \label{c5.2} \\ 
& \pi^{sp}_{kt} - \pi^{sp}_{k(t+1)} - \pi^{wo}_{k(t+1)} + \pi^{sf}_{kt} - \pi^{sf}_{k(t+1)} \leq 0, \qquad   k\in K, \enspace   t\in \{1,..,T-1\} \label{c5.3} \\ 
& \pi^{sp}_{kT} - \pi^{ch}_{k} \leq 0, \qquad   k\in K \label{c5.4} \\
& \pi^{sp}_{kt} + \pi^{wo}_{kt} \leq 1, \qquad   k\in K, \enspace   t\in \{1,..,T-1\} \label{c5.5} \\
& \pi^{sp}_{kT} + \pi^{wo}_{kT} + \pi^{cf}_{k} \leq 1, \qquad   k\in K \label{c5.8} \\
& \pi^{sp}_{kt}, \pi^{wo}_{kt}, \pi^{sf}_{kt}, \pi^{cf}_{k} \geq 0, \qquad    k\in K, \enspace   t\in T \label{c5.6} \\
& \pi^{fs}_{k}, \pi^{ch}_{k} \text{unrestricted}, \qquad   k\in K \label{c5.7} 
\end{flalign}
\end{subequations}

We provide improvements to the dual problem in several ways. The dual variables $\pi_{fs}$ and $\pi_{cf}$ are removed since the corresponding subproblem constraints \eqref{c3.7} and \eqref{c3.8} are eliminated in the improved model. The dual variables $\pi^{fs}$ and $\pi^{ch}$ are not in the objective function and are unrestricted, which means that we can remove these variables and the constraints with those variables from the formulation without altering the optimal value of the problem. In our preliminary computational studies, we improved the dual subproblem by removing these variables. However, we observed that most of the DSPs have multiple optimal solutions, and as the number of vehicles and stations increase, it is more likely to have multiple optimal solutions. This naturally raises the question of what optimal dual vector provides the strongest cut, if we add only one cut per iteration per scenario. One can potentially add the cuts corresponding to all optimal dual extreme points, however, this results in an explosion in the size of the relaxed master problem after just a couple of iterations. While there is no reliable way to identify the weak cuts \citep{rahmaniani2017benders},  we executed experiments in order to find a pattern for strong cuts. Our findings showed that adding the cut corresponding to the optimal dual extreme point with the most non-zero variables results in the fastest convergence. Thus, we added an $\ell_1$ regularization term to the objective function of the DSP, hence, the new objective is encouraged to choose an optimal solution with the most non-zero variables. Accordingly, we propose an improved DSP formulation as follows: 
\begin{subequations}
\label{form: dual-improved}
\begin{flalign}
 \max_{\pi} \quad & \sum_{k\in K}\sum_{t\in T} \pi^{sp}_{kt} (\sum_{v\in V} p_{kv}\hat{x}_{vt}-c+\epsilon) + \pi^{wo}_{kt} (\sum_{v\in V} p_{kv}\hat{x}_{vt}-l_k+\epsilon) \label{c6.0} \\
\text{s.t.} \quad & \pi^{sp}_{kt} - \pi^{sp}_{k(t+1)} -\pi^{wo}_{k(t+1)}  \leq 0, \qquad   k\in K, \enspace   t \in \{1,..,T-1\} \label{c6.1} \\ 
& \pi^{sp}_{kt} + \pi^{wo}_{kt} \leq 1, \qquad   k\in K, \enspace   t\in T \label{c6.2} \\
& \pi^{sp}_{kt}, \pi^{wo}_{kt} \geq 0, \qquad   k\in K, \enspace   t\in T \label{c6.3} 
\end{flalign}
\end{subequations}

\subsection{Heuristic Solution Approach}\label{section: mms-heuristics}
MMS is an NP-hard problem, and stochastic failures of products (cars) increases the computational burden of solving  the problem drastically. Hence, it is essential to create efficient heuristic procedures in order to solve industry-sized problems. In this section, we provide a fast and easy-to-implement greedy heuristic to find a good initial feasible first-stage decision (i.e., a sequence of vehicles) and an efficient tabu search (TS) algorithm to improve the solution quality. Although all $|V|!$ vehicle permutations are  feasible, the proposed greedy heuristic aims to find a good initial feasible solution. To achieve this, a solution is generated for the  deterministic counterpart of the proposed MMS problem, which excludes vehicle failures. We refer to this problem as the \textit{ one-scenario problem}, since the corresponding problem has a single scenario with no failed vehicles. Assuming that the failure probability of each vehicle is less than or equal to 0.5, the scenario with no failed vehicles has the highest probability. Once such a feasible sequence of vehicles is generated, the TS  algorithm improves this solution in two parts: first, over the one-scenario problem, and then, over the full-information problem.

\subsubsection{Greedy Heuristic}\label{section: mms-greedy-heuristic}
It is important for a local search heuristic algorithm to start with a good quality solution. A naive approach to generate an initial solution (sequence) is to  always select the vehicle that causes the minimum new work overload for the next position. However, this approach is myopic since it only considers the current position. We remediate this issue by decreasing future work overloads which includes considering idle times and dynamic utilization rates. Accordingly, in order to generate a good initial solution, we propose utilizing an iterative greedy heuristic that follows a priority rule based on the work overload, idle time, and weighted sum of processing time, to be defined shortly. 

Before explaining our proposed greedy heuristic, let us define some technical terms. The \textit{idle time} refers to the duration that an operator waits for the next vehicle to enter the station borders. The weight of processing times is determined by using station \textit{utilization rates}, which is inspired by car sequencing problem utilization rates \citep{solnon2000solving, gottlieb2003study}. We describe the utilization rate of a station as the ratio between the average processing time on a station and the cycle time, so the utilization rate of station $k$ is $\sum_{v\in V}p_{kv} / (|V|*c)$. At each iteration, after a new assignment of a vehicle, the dynamic utilization rates are calculated by considering only the unassigned vehicles. Accordingly, the weighted sum of the processing time of a vehicle $v$ is calculated using \eqref{eq: weighted-sum processing time}:
\begin{equation}\label{eq: weighted-sum processing time}
    \frac{\sum_{k\in K}p_{kv}\sum_{i\in \hat{V}}p_{ki}}{|K|*|\hat{V}|*c},
\end{equation}
where $\hat{V}$ denotes the set of unassigned vehicles. 
If the utilization rate of a station is greater than 1, then the average processing time is more than the cycle time, which induces an unavoidable work overload. On the other hand, a utilization rate close to 0 indicates that the average processing time is minimal compared to the station's allocated time. 

Our proposed greedy heuristic builds a sequence iteratively, one position at a time, starting from the first position and iterating over positions. We use $t$ to denote an iteration. At each iteration, $t={1,\ldots,T}$, the unassigned vehicles that cause the minimum new work overload are determined, denoted by $V_{t,wo}$. Ties are broken by selecting the vehicles from the set $V_{t,wo}$, which causes the minimum new idle time, the new set of vehicles is denoted by $V_{t,idle}$. In the case of ties, the vehicle with the highest weighted sum of processing time from the set $V_{t,idle}$ is assigned to the position $t$ of the sequence. Note that the first vehicle of the sequence is the vehicle with the highest weighted sum of processing time among the set $V_{0,idle}$ since there is no work overload initially.

Finally, we enhance the proposed greedy heuristic by considering the category of the vehicles. Motivated by our case study, we categorize the vehicles  based on the engine type, electric or non-electric, because  the engine type is the most restrictive feature due to the high EV ratio (number of EVs divided by the number of all vehicles). Moreover, the engine type leads to  different processing times  on a specific station. Hence, we modify our greedy heuristic to first decide whether an EV or a non-EV should be assigned to the next position at each iteration. Accordingly, first, the EV ratio is calculated, and  an EV is assigned to the first position. The procedure always follows the EV ratio. For example, if the EV ratio is $1/3$, an EV will be assigned to the positions $1+3t$ where $t=\{1,\ldots,\frac{|T|}{3}-1\}$. In case of a non-integer EV ratio, the position difference between any two consecutive EVs is the integer part of the ratio plus zero or one; randomly decided based on the decimal part of the ratio.
Once the vehicle category is decided throughout the entire sequence, the specific vehicle to be assigned is selected based on the above-described procedure. 
We note that this enhancement in the greedy heuristic may be applied for any restrictive feature that causes large variations in  processing times.

\begin{table}[h]
\centering
\scalebox{0.8}{
\begin{tabular}{
>{\columncolor[HTML]{FFFFFF}}c 
>{\columncolor[HTML]{FFFFFF}}c 
>{\columncolor[HTML]{FFFFFF}}c 
>{\columncolor[HTML]{FFFFFF}}c l}
\multicolumn{1}{l}{\cellcolor[HTML]{FFFFFF}Vehicle} & \multicolumn{1}{l}{\cellcolor[HTML]{FFFFFF}Engine} & $p_1$ & $p_2$ &  \\ \cline{1-4}
A                                                            & Electric                                                    & 15            & 4    &  \\
B                                                            & Electric                                                    & 16            & 3    &  \\
C                                                            & Gasoline                                                    & 2             & 10   &  \\
D                                                            & Gasoline                                                    & 3             & 8    &  \\
E                                                            & Gasoline                                                    & 2             & 9    &  \\
F                                                            & Gasoline                                                    & 4             & 7    & 
\end{tabular}}
\caption{Illustration of greedy heuristic}
\label{tbl: greedy-heuristic-example}
\end{table}

To describe the greedy heuristic, consider an example with six vehicles and two stations. The processing times and engine types of vehicles are given in Table \ref{tbl: greedy-heuristic-example}. The cycle time is 7 TU, and the length of the stations are 20 TU and 10 TU, respectively. The EV ratio is 1/3. We consider only EVs for the first position, vehicle A is designated to the first position since it causes less idle time than vehicle B. Next, none of the non-EVs causes work overload or idle time, so we assign the vehicle with the highest weighted sum of processing times to the second position, vehicle C. The procedure continues with another non-EV, and vehicle F is assigned to the third position because it is the only vehicle that does not cause any work overload. Consistent with the $1/3$ EV ratio, an EV must be assigned to the fourth position, and vehicle B is assigned to this position as it is the only EV left. Vehicle E is assigned to the fifth position due to its higher weighted sum of processing times. Finally, vehicle D is assigned to the last position. The resulting sequence is \textit{A-C-F-B-E-D} with a work overload of 3 TU, only at position 6 at station 2.

\subsubsection{Tabu Search Algorithm}\label{section: mms-tabu-search}
This section proposes a simulation-based local search algorithm on a very large neighborhood with tabu rules. The TS algorithm starts at the initial feasible solution (sequence) generated by the iterative greedy heuristic, and improves the initial solution via iterative improvements within the designed neighborhood. At each iteration of the TS, a transformation operator is randomly selected based on operator weights and applied to the incumbent solution to visit a random neighbor, respecting the tabu rules. The candidate solution is accepted if the objective function value is non-deteriorated, i.e., the candidate solution is rejected only if it has more total work overload. Then, another random operator is applied to the incumbent solution. This process repeats until the stopping criterion is met. 

As aforementioned, the TS has two parts. The first part acts as the second step of the initial solution generation procedure since it improves the solution provided by the greedy heuristic for the one-scenario problem. In our preliminary numerical experiments we observed that this step can drastically improve the initial solution quality. Hence, we conduct this step for a duration $\tau_{one}$. Next, the algorithm makes a transition to the full-information problem and reevaluates the objective function value of the incumbent solution---the sequence generated by the first part of TS. In the second part of the TS algorithm, the objective function value corresponding to the sequence is evaluated for the full-information problem. 
To do this, we calculate the total work overload for all realizations $\omega \in \Omega$, given the first-stage decision (sequence). That is, we calculate the objective function of  \eqref{form: 2nd_standard} for each  realization $\omega \in \Omega$ and take the weighted sum, each multiplied by the probability of the scenario. Observe from \eqref{form: 2nd_standard} that once the first-stage decision is fixed, the problem decomposes in scenarios and stations. Accordingly, the solution evaluation process is parallelized over scenarios and stations. 

The TS algorithm continues evaluating the solution for the full-information problem for a duration  $\tau_{full}$. The allocated time for the second part, $\tau_{full}$, is much larger than that of the first part, $\tau_{one}$, since iterating over one-scenario  problem is much faster than that over a set of realizations. 
In the reminder of this section, we explain various components of the TS algorithm. 

\paragraph{Objective Evaluation}
The objective function of the problem for a given scenario is the same as the objective given in \eqref{c3.0}, total work overload over all stations and positions. Evaluation of the objective, after each movement, is the bottleneck of our algorithm since the new total work overload needs to be determined. Note that the objective evaluation starts at the first position and is executed iteratively since there is a sequence dependency. Accordingly, we propose to reduce the computational burden in two ways. 

First, reevaluating the whole sequence is unnecessary since transformation operators make local changes in the sequence, i.e., some parts of the sequence remain unaltered and do not require reevaluation.
Hence, we apply partial reevaluation after each movement. To explain partial reevaluation, assume that the vehicles at positions $t_1$ and $t_2$ are swapped. We certainly know that the subsequence corresponding to positions $[1, t_1-1]$ is not impacted by the swap operation; hence, we do not reevaluate these positions. Additionally, we may not have to reevaluate all the positions in $[t_1, t_2-1]$ and the positions in $[t_2, |T|]$. In each of these subsequences, there could be a \textit{reset position} which ensures that there is no change in the objective from that position until the end of the subsequence. Since the rest of the subsequence after the reset position is not changed, we can jump to the end of the subsequence. To highlight how a partial reevaluation may improve the objective reevaluation process, suppose that the vehicles at positions 350 and 380 are swapped. We certainly know the subsequence corresponding to positions [1, 349] is not impacted by the swap. Additionally, in the case that there is a reset point before position 380 (and $|T|$), we do not have to reevaluate all the positions between 350 and 380, and the positions between 380 and $|T|$. 
 
Second, we calculate the objective function in an accelerated way. Traditionally work overload and starting position for position $t$ at station $k$, respectively $w_{kt}$ and $z_{k(t+1)}$, are calculated as: $w_{kt}=z_{kt}+b_{kt}-l_{k}$ and $z_{k(t+1)}= z_{kt}+b_{kt}-w_{kt}-c$,  where $w_{kt}, z_{kt} \geq 0$. Instead of calculating work overload and starting position vectors separately, we propose using a single vector to extract these information, which in fact is a different representation of the starting position vector $z$. If there is a work overload at position $t$, then $z_{k(t+1)}=l_{k}-c$. Otherwise, if there is not any work overload at position $t$, then $z_{k(t+1)}=z_{kt}+b_{kt}-c$, or we can equivalently write $z_{k(t+1)}=z_{k(t-1)}+b_{k(t-1)}-c+b_{kt}-c-w_{k(t-1)}$. Again, if there is a work overload at position $t-1$, then $z_{k(t+1)}=(l_{k}-c)+b_{kt}-c$, otherwise, if there is not any work overload at $t-1$, then $z_{k(t+1)}=z_{k(t-1)}+(b_{k(t-1)}-c)+(b_{kt}-c)$. Since we know that $z_{k0}=0$, we can generalize it as $z_{k(t+1)}=\sum_{h=1}^{t}(b_{kh}-c)$, which is the cumulative sum of vector $\eta_{k}=b_{k}-c$ up to and including position $t$. However, this generalization has the assumption that there is not any work overload or idle time up to position $t$. We note that there is an idle time at position $t+1$ when $z_{kt}+b_{kt}-c < 0$. Accordingly, we can write a general formula as $z_{k(t+1)}=\max(0, \min(l_{k}-c, z_{kt}+\eta_{k(t+1)}))$,  which is referred to as the conditional cumulative sum of $\eta_{k}$ up to position $t$. Intuitively, the conditional cumulative sum is defined as follows: starting from position 0, the cumulative sum is calculated iteratively within the closed range $[0,l_{k}-c]$. Whenever the cumulative sum exceeds the lower bound zero or the upper bound $l_{k}-c$, we set the cumulative sum to the corresponding bound's value. If the cumulative sum is below the lower bound, the excess value is equal to the idle time. Otherwise, if the  cumulative sum  is above the upper bound, the excess value is equal to the work overload. For example,  if the cumulative sum is -2 at a position, the cumulative sum is set to zero and there is a 2 TU of idle time at that position.

In light of the proposed improvements, the partial reevaluation process is executed in two subsequences, $[t_1, t_2)$ and $[t_2, |T|]$, assuming that $t_1$ and $t_2$ are the two selected positions by any transformation operator and $t_1 < t_2$. The process starts at the first position, called position one, of the corresponding subsequence. We set $z_{k0}=\eta_{k1}$ and we calculate the starting position, work overload, and idle time for the positions in the subsequence iteratively as mentioned above. The reevaluation of the subsequence is completed when either a reset position is found or the whole subsequence is iterated. A reset position occurs at position $t$ differently in two cases as follows: 1) if $z_{k,t+1}=0$ when the processing time on the starting position $t_1$ (or $t_2$) is decreased, 2) if the sum of idle time and work overload up to position $t$ in the current subsequence exceeds the total increased processing time at the corresponding starting position $t_1$ (or $t_2$) when the processing time on the starting position $t_1$ (or $t_2$) is increased. 

\paragraph{Transformation Operators}
In this section, we explain the details of the transformation operators. We employ swap, forward and backward insertions, and inversion operators. The swap operator interchanges the positions of two randomly selected cars. Insertion removes a car from position $i$ and inserts it at position $j$. Insertion is applied in two different directions, backward and forward. When $i>j$, the insertion is called a backward insertion, and all the vehicles between the positions $j$ and $i$ move one position to the right , i.e., scheduled later. On the contrary, forward insertion occurs when $i<j$ and all the vehicles between the positions $i$ and $j$ move one position to the left, i.e., scheduled earlier. Inversion takes two randomly selected positions in the sequence, and the subsequence between the selected positions is reversed. A repetitive application of these operators creates a very large neighborhood which helps the improvement procedure to escape local optima, especially when it is combined with a non-deteriorated solution acceptance procedure. The latter enables the algorithm to move on the plateaus that consist of the solutions with the same objective function value (see Section \ref{sec: compu} for numerical experiments). 

\paragraph{Tabu List}
We design the tabu list in a non-traditional manner. This list includes the movements that induce undesired subsequences. Based on our observations, we define an undesired subsequence as back-to-back EVs because consecutive EVs cause a tremendous amount of work overload at the battery loading station. Accordingly, any movement that results in back-to-back EVs is a tabu. For the sake of clarity, we describe tabu movements in detail for each operator separately in \ref{appendix: mms-tabu-list}.

\subsection{SAA Approach and Solution Quality Assessment}\label{mms: solution quality assessment}
In \eqref{eq:true stochastic problem}, it is assumed that the probability of each scenario is known a priori, which may not hold in practice. In addition, the exponential growth of the number of scenarios causes an explosion in the size of the stochastic program. Hence, we utilize the SAA approach to tackle these issues. Consider the abstract formulation \eqref{eq:true stochastic problem}. The SAA method approximates the expected value function with an identically and independently distributed (i.i.d) random sample of $N$ realizations of the random vector $\Omega_{N}:=\{\omega_1,\ldots,\omega_N\}\subset \Omega$ as follows:
\begin{flalign}\label{eq: generic SAA problem}
& z_N= \min_{x \in X} \frac{1}{N} \sum_{\omega\in \Omega_{N}} Q(x,\xi_\omega).
\end{flalign}
The optimal value of \eqref{eq: generic SAA problem}, $z_N$, provides an estimate of the true optimal value \citep{kleywegt2002sample}. Let $\hat{x}_N$ and $x^*$ denote an optimal solution to the SAA problem \eqref{eq: generic SAA problem} and the true stochastic program \eqref{eq:true stochastic problem},  respectively.
Note that $\mathbb{E}[Q(\hat{x}_N,\xi_\omega)] - \mathbb{E}[Q(x^*,\xi_\omega)] $ is the optimality gap of solution $\hat{x}_N$, where $\mathbb{E}[Q(\hat{x}_N,\xi_\omega)]$ is the (true) expected cost of solution $\hat{x}_N$ and $\mathbb{E}[Q(x^*,\xi_\omega)]$ is the optimal value of the true problem \eqref{eq:true stochastic problem}. A high quality solution is implied by a small optimality gap. However, as $x^*$ (and hence, the optimal value  of the true problem) may not be known, one may obtain a statistical estimate of the optimality gap to assess the quality of the candidate solution $\hat{x}_N$ \citep{homem2014monte}. That is, given that $\mathbb{E}[z_N] \le \mathbb{E}[Q(x^*,\xi_\omega)]$, we can obtain an upper bound on the optimality gap as $\mathbb{E}[Q(\hat{x}_N,\xi_\omega)] - \mathbb{E}[z_N]$. 
We employ the multiple replication procedure of \citet{mak1999monte} in order to assess the quality of a candidate solution by estimating an upper bound on its optimality gap. A pseudo-code for this procedure is given in Algorithm \ref{pseudo-code:MRP}. 
We utilize MRP, in Section \ref{section: mms-computational-experiments}, to assess the quality of solutions  generated by different approaches.

\begin{algorithm}[h]
\scriptsize
\caption{Multiple Replication Procedure $\textrm{MRP}_{\alpha}(\hat{x})$ }\label{pseudo-code:MRP}
\begin{algorithmic}
\STATE{\textbf{Input:} Candidate solution $\hat{x}$, replication size $M$, and $\alpha \in (0,1)$.}
\STATE{\textbf{Output:} A normalized $\%100(1-\alpha)$ upper bound on the optimality gap of $\hat{x}$.}
\FOR{$m= 1,2,\ldots, M$.}
    \STATE {Draw i.i.d. sample $\Omega^{m}_N$ of realizations $\xi^{m}_\omega$, $\omega \in \Omega_{N}^{m}$.}
    \STATE {Obtain $z^m_N := \min\limits_{x \in X} \frac{1}{N} \sum\limits_{\omega \in \Omega_{N}^{m}} Q(x,\xi^{m}_\omega)$.}
    \STATE{Estimate the out-of-sample cost  of $\hat{x}$ as      
             $\hat{z}^m_N := \frac{1}{N}\sum\limits_{\omega \in  \Omega_{N}^{m}}Q(\hat{x},\xi^{m}_\omega)$.}
    \STATE {Estimate the optimality gap of $\hat{x}$ as $G^m_N := \hat{z}^m_N - z^m_N$.}
\ENDFOR
\STATE {Calculate the sample mean and sample variance of the gap as }
\STATE {$\qquad \Bar{G}_{N}=\frac{1}{M}\sum_{m=1}^{M}G_{N}^{m}\quad$ and $\quad s_{G}^{2}=\frac{1}{M-1}\sum_{m=1}^{M}(G_{N}^{m}-\Bar{G}_{N})^{2}$.}
\STATE {Calculate a normalized $\%100(1-\alpha)$ upper bound on the optimality gap as}
\STATE{$\frac{1}{\bar{z}_N}\left(\bar{G}_N + t_{\alpha;M-1}\frac{s_G}{\sqrt{M}}\right)$, where  $\bar{z}_{N}=\frac{1}{M} \sum_{m=1}^{M} z^m_N $.}
\end{algorithmic}
\end{algorithm}
In addition, we propose an MRP integrated SAA approach for candidate solution generation and quality assessment, given in Algorithm \ref{pseudo-code:MRP-integrated-SAA}. On the one hand, a candidate solution is generated by solving an SAA problem with a sample of $N$ realizations. Then, we use the MRP to estimate an upper bound on the optimality gap of the candidate solution. If the solution is $\epsilon$-optimal, i.e.,  estimated upper bound on its optimality gap is less than or equal to a $\epsilon$ threshold, the algorithm stops. Otherwise, the sample size increases until a good quality solution is found. 
The algorithm returns a candidate solution and its optimality gap. 

\begin{algorithm}[h]
\scriptsize
\caption{MRP integrated SAA}\label{pseudo-code:MRP-integrated-SAA}
\begin{algorithmic}
\STATE{\textbf{Input:} List of sample sizes $N_{list}$ and  $\epsilon,\alpha \in (0,1)$.}
\STATE{\textbf{Output:} Solution $\hat{x}$ and $\textrm{OptGap}$.}
\FOR{$N$ in $N_{list}$}
    \STATE {Obtain a candidate solution $\hat{x}_N$ by solving the SAA problem \eqref{eq: generic SAA problem}.}
    \STATE{Calculate a normalized $\%100(1-\alpha)$ upper bound on the optimality gap as $\textrm{MRP}_{\alpha}(\hat{x}_N)$.}
    \IF{$\textrm{MRP}_{\alpha}(\hat{x}_N) \leq \epsilon$} 
        \STATE{$\hat{x} \leftarrow \hat{x}_N$ and $\textrm{OptGap} \leftarrow \textrm{MRP}_{\alpha}(\hat{x}_N)$.}
        \STATE {\textbf{exit for loop}}
    \ENDIF
\ENDFOR

\end{algorithmic}
\end{algorithm}

We end this section by noting that each of the DEF, presented in Section \ref{section: mms-mathematical-model}, the L-shaped algorithm, presented in Section \ref{section: mms-exact-solution}, and the heuristic algorithm, presented in Section \ref{section: mms-heuristics}, can be used to solve the SAA problem and obtain a candidate solution. However, the probability of scenarios $\omega \in \Omega$, $\rho_{\omega}$, must  change in the formulations so that it reflect the scenarios in a sample $\Omega_N$.  
Let $\hat{N}$ and $n_{\omega}$ represent the set of unique scenarios in $\Omega_N$ and the number of their occurrences. Thus, in the described DEF, L-shaped algorithm, and TS algorithm, $\sum_{\omega \in \Omega} \rho_\omega (\cdot)$ changes to $\frac{1}{N}\sum_{\omega \in \Omega_{N}} (\cdot)$ or equivalently, $\frac{1}{N}\sum_{\omega \in \hat{N}} n_\omega(\cdot)$. 
Accordingly, in the L-shaped method, we generate one optimality cut for each unique scenario $\omega\in \hat{N}$ by solving $|\hat{N}|$ number of subproblems at each iteration.

\section{Numerical Experiments}\label{section: mms-computational-experiments}
In Section \ref{section: mms-experimental-setup}, we describe the experimental setup. Then, in Section \ref{section: mms-exact-results} and \ref{section: mms-heuristic-results}, we assess solution quality and computational performance of the proposed L-shaped and heuristic algorithms applied to a SAA problem, respectively. 

\subsection{Experimental Setup}\label{section: mms-experimental-setup}
We generated real-world inspired instances from our automobile manufacturer partner's assembly line and planning information. As given in Table \ref{tbl: mms-dataset table}, we generated three types of instances: (1) small-sized instances with 7-10 vehicles to assess the performance of L-shaped algorithm, (2) medium-sized instances with  40 vehicles to assess the performance of the TS algorithm for the one-scenario problem, (3) large-sized instances with 200, 300, and 400 vehicles to evaluate the performance of the TS algorithm. All instances have five stations, of which the first one is selected as the most restrictive station for EVs, the battery loading station. The rest are selected among other critical stations that conflict with the battery loading station. 
\begin{table}[h]
\centering
\scalebox{0.8}{
\begin{tabular}{
>{\columncolor[HTML]{FFFFFF}}l 
>{\columncolor[HTML]{FFFFFF}}c 
>{\columncolor[HTML]{FFFFFF}}c 
>{\columncolor[HTML]{FFFFFF}}c }
 Instance Type                           & $|V|$ & $|K|$ & Number of Instances \\ \hline
Small   & 7, 8, 9, 10                   & 5                 & 30 $\times$ 4               \\
Medium  & 40                        & 5                 & 30 $\times$ 1               \\
Large    & 200, 300, 400                & 5                 & 30 $\times $ 3             
\end{tabular}}
\caption{Data sets}
\label{tbl: mms-dataset table}
\end{table}
The cycle time $c$ is 97 TU, and the station length $l$ is 120 TU for all but the battery loading station, which is two station lengths, 240 TU. The information about the distribution of the processing times is given in Table \ref{tbl:mms-processing-times}. It can be observed that the average and maximum processing times for each station are lower than the cycle time and the station length, respectively. Moreover, the ratio of the EVs is in the range of [0.25, 0.33] across all instances. 

\begin{table}[h]
\centering
\scalebox{0.8}{
\begin{tabular}{
>{\columncolor[HTML]{FFFFFF}}c 
>{\columncolor[HTML]{FFFFFF}}c 
>{\columncolor[HTML]{FFFFFF}}c 
>{\columncolor[HTML]{FFFFFF}}c}
\multicolumn{1}{l}{\cellcolor[HTML]{FFFFFF}}           & \multicolumn{3}{c}{\cellcolor[HTML]{FFFFFF}Time (s)} \\ \cline{2-4} 
\multicolumn{1}{l}{\cellcolor[HTML]{FFFFFF}Station ID} & Min             & Mean           & Max  \\  \hline  
1                                                      & 42.6            & 94.1                          & 117.2           \\
2                                                      & 7.9             & 84.3                          & 197.9           \\
3                                                      & 57.8            & 96.2                          & 113.3           \\
4                                                      & 26.9            & 96.9                          & 109.7           \\
5                                                      & 57.8            & 96.2                          & 114.3          
\end{tabular}}
\caption{Processing times distribution}
\label{tbl:mms-processing-times}
\end{table}

We derived the failure rates from six months of historical data by performing predictive feature analysis on vehicles. Based on the analysis, two groups of vehicles are formed according to their failure probabilities, low-risk and high-risk vehicles, whose failure probabilities are in the range of [0.0, 0.01] and [0.2, 0.35], respectively. The failure probability is mostly higher for recently introduced features, e.g., the average failure probability of EVs is 50\% higher than that of other vehicles. High-risk vehicles constitute  [0.03, 0.05] of all vehicles. However, this percentage increases to [0.15, 0.25] for the small-sized instances in order to have a higher rate of failed vehicles. We note that the failures are not considered for the medium-sized instances since these instances are used for only the one-scenario problem, which does not involve failures by definition. 

The number of failure scenarios, $2^{|V|}$, increases exponentially in the number of vehicles. Thus, we generated an i.i.d random sample of $N$ realizations of the failure scenarios; hence, formed a SAA problem. For each failure scenario and vehicle, we first chose whether the vehicle was high risk or low risk (based on their prevalence). Then, depending on being a high-risk or low-risk vehicle, a failure probability was randomly selected from the respective range. Finally, it was determined whether the vehicle failed or not. In order to have a more representative sample of scenarios for large-sized instances, no low-risk vehicle was allowed to fail at any scenario.

For each parameter configuration, we generated 30 instances. The vehicles of each instance were randomly selected  from a production day, respecting the ratios mentioned above. The algorithms were implemented in Python 3. For solving optimization problems we used Gurobi 9.0. The time limit is 600 seconds for all experiments unless otherwise stated. We run our experiments on computing nodes of the Clemson University supercomputer. The experiments with the exact solution approach were run on nodes with a single core and 15 GB of memory, and the experiments with the heuristic solution approach were on on nodes with 16 cores and 125 GB of memory.

\subsection{Exact Solution Approach}\label{section: mms-exact-results}
In this section, we present results for the solution quality and computational performance of the L-shaped algorithm. We used MRP scheme,  explained in Section \ref{mms: solution quality assessment}, to assess the solution quality. We also compared the computational performance of the L-shaped algorithm with that of solving the DEF. We present the results for 120 small-sized instances consisting of 7 to 10 vehicles. We do not present the results for large-sized instances as our preliminary experiments showed that the number of instances that could be solved to optimality decreases drastically. 

We also point that instead of solving a relaxed master problem to optimality at each iteration of the L-shaped algorithm, one can aim for just obtaining a feasible solution $\hat{x} \in X$. This may result in saving a significant amount of computational time that would be otherwise spent on exploring solutions that are already eliminated in previous iterations. This kind of implementation, referred to as {\it branch-and-Benders-cut} (B\&BC), is studied in the literature, see, e.g., \citep{hooker2011logic, thorsteinsson2001branch, codato2006combinatorial}. In our implementation of our proposed L-shaped algorithm, we used Gurobi's Lazy constraint callback to generate cuts at a feasible integer solution found in the course of the branch-and-bound algorithm. 

\subsubsection{Solution Quality}\label{sec: exact_qual}
\begin{figure}[h]
\captionsetup{justification=centering}
    \centerline{\includegraphics[scale=0.7]{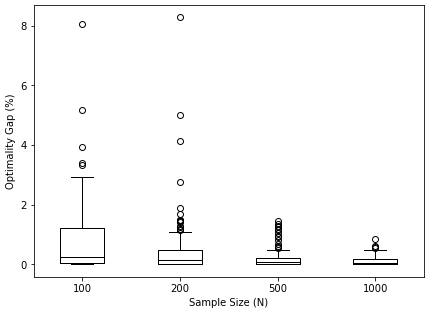}}
    \caption{Solution quality of the SAA problem based on sample sizes}
    \label{fgr:solution_quality_exact}
\end{figure}

Figure \ref{fgr:solution_quality_exact} shows the impact of sample size on the solution quality of the SAA problem. Observe that the improvement in the upper bound of the optimality gap, the MRP output, as the sample size increases from 100 to 1000 progressively. We set the number of replications $M$ to 30 and $\alpha=0.05$ (95\% confidence interval). While the mean of the optimality gap decreases gradually from 0.76\% to 0.12\%, a drastic enhancement is observed with the variance. We have 36 out of 120 solutions with an optimality gap of larger than 1\% when the sample size is 100. However, all of the obtained solutions have less than a 1\% optimality gap when the sample size is 1000. It can be seen in the figure that good solutions can be obtained with a sample size of 100, yet it is not assured due to the high variance of the approximation. Consequently, the results suggest that the sample size should be increased until the variance of objective estimation is small enough.

Based on the results in Figure \ref{fgr:solution_quality_exact}, we implemented the MRP integrated SAA scheme, presented in Section \ref{mms: solution quality assessment} and Algorithm \ref{pseudo-code:MRP-integrated-SAA}, to balance the required computational effort and solution quality. We set $\alpha=0.05$ (95\% confidence interval) and $\epsilon=0.01$ in MRP. While it is ensured that we obtain solutions within a 1\% optimality gap, most of the solutions are found with the least computational effort, e.g., at the first iteration with a sample size of 100. In Table \ref{tbl: mms-saa performance}, we provide key performance results to show the performance of the MRP integrated SAA scheme, where the number of replications $M$ is 30 and the MRP sample size $N$ is 5000. The average value for the optimal value, accepted candidate solution's expected objective value, and optimality gap are presented in Table \ref{tbl: mms-saa performance}. The sample size of the accepted candidate solutions is 84, 20, 11, and 5 for the sample sizes $N_{list}=\{100, 200, 500, 1000\}$, respectively. The average optimality gap is 0.2\% which shows that SAA can produce high-quality solutions.
\begin{table}[h]
\centering
\scalebox{0.8}{
\begin{tabular}{lllll}
\cellcolor[HTML]{FFFFFF}\begin{tabular}[c]{@{}l@{}}Statistical   Lower \\      Bound $\mathbb{E}[z_{N}]$ \end{tabular} & \cellcolor[HTML]{FFFFFF}\begin{tabular}[c]{@{}l@{}}Estimated Objective\\      Value $\mathbb{E}[Q(\hat{\textit{x}}_{N}, \xi_{\omega})]$ \end{tabular} & \cellcolor[HTML]{FFFFFF}\begin{tabular}[c]{@{}l@{}}Optimality \\      Gap $\Bar{G}_{N}$\end{tabular} &  &  \\ \cline{1-3}
\cellcolor[HTML]{FFFFFF}55.80    & \cellcolor[HTML]{FFFFFF}55.91       & \cellcolor[HTML]{FFFFFF}0.11 (0.2\%)         
\end{tabular}}
\caption{Solution quality of the MRP integrated SAA}
\label{tbl: mms-saa performance}
\end{table}

\begin{figure}[h]
\captionsetup{justification=centering}
    \centerline{\includegraphics[scale=0.7]{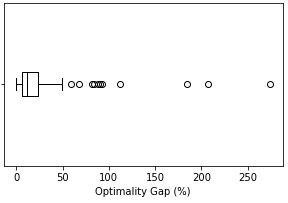}}
    \caption{Solution quality of the one-scenario problem}
    \label{fgr:deterministic_solution_quality_exact}
\end{figure}

Additionally, we assess the solution quality of the one-scenario problem (i.e., the deterministic MMS problem without any car failure). Observe from Figure \ref{fgr:deterministic_solution_quality_exact} that the average optimality gap is 23\%, the maximum optimality gap is 274\%, and the standard deviation is 39\%. Comparing the performance of the SAA and the one-scenario problems shows that we can generate robust solutions by considering vehicle failures, which helps  reduce work overloads by more than 20\%.

\subsubsection{Computational Performance}
\label{sec: exact_comp_per}
In this section, we conduct different experiments to compare the DEF and L-shaped algorithm. On the one hand, we assess the impact of using the {\it improved} model, described in Section \ref{section: mms-mathematical-model}, obtained by setting the processing time of failed vehicles to the cycle time, and compare the results with those obtained using the standard model.  The DEF corresponding to the standard and improved models are denoted as $D_{std}$ and $D_{imp}$, respectively. Similarly, the L-shaped algorithm corresponding to the standard and improved are denoted as $L_{std}$ and $L_{imp}$. 
On the other hand, we assess the impact of our proposed cut selection strategy, described in Section \ref{section: mms-exact-solution}, obtained using $\ell_1$ norm regularization to find a cut with the least number of zero coefficients. We used the cut selection strategy for the improved model, and denote the corresponding L-shaped algorithm as $L_{imp-cs}$. 

In Table \ref{tbl: mms-exact-solution-results table}, we  present the results on the impact of the improved model and cut selection strategy to compare the DEF and L-shaped algorithm for solving the SAA problem. We report the average and standard deviation of the computational time (in seconds), labelled as $\mu_t$ and $\sigma_t$, respectively, and the optimality gap, labelled as \textit{Gap} (in percentage). Additionally, the number of instances that could not be solved optimally within the time limit is given in parenthesis under the \textit{Gap} columns. All time results are the average of instances (out of 30 instances) that could be solved optimally within the time limit, while the \textit{Gap} results are the average of the instances that could not be solved optimally. Based on the results in Section \ref{sec: exact_qual}, we conducted the computational experiments on the SAA problem with sample sizes 100, 200, 500, and 1000.

\begin{table}[h]
\centering
\resizebox{\columnwidth}{!}{%
\begin{tabular}{
>{\columncolor[HTML]{FFFFFF}}r 
>{\columncolor[HTML]{FFFFFF}}c 
>{\columncolor[HTML]{FFFFFF}}c 
>{\columncolor[HTML]{FFFFFF}}c 
>{\columncolor[HTML]{FFFFFF}}c l
>{\columncolor[HTML]{FFFFFF}}c 
>{\columncolor[HTML]{FFFFFF}}c 
>{\columncolor[HTML]{FFFFFF}}c l
>{\columncolor[HTML]{FFFFFF}}c 
>{\columncolor[HTML]{FFFFFF}}c 
>{\columncolor[HTML]{FFFFFF}}c l
>{\columncolor[HTML]{FFFFFF}}c 
>{\columncolor[HTML]{FFFFFF}}c 
>{\columncolor[HTML]{FFFFFF}}c l
>{\columncolor[HTML]{FFFFFF}}c 
>{\columncolor[HTML]{FFFFFF}}c 
>{\columncolor[HTML]{FFFFFF}}c }
\multicolumn{1}{c}{\cellcolor[HTML]{FFFFFF}} &   & \multicolumn{3}{c}{\cellcolor[HTML]{FFFFFF}$D_{std}$} & & \multicolumn{3}{c}{\cellcolor[HTML]{FFFFFF}$D_{imp}$} & & \multicolumn{3}{c}{\cellcolor[HTML]{FFFFFF}$L_{std}$} & & \multicolumn{3}{c}{\cellcolor[HTML]{FFFFFF}$L_{imp}$} & & \multicolumn{3}{c}{\cellcolor[HTML]{FFFFFF}$L_{imp-cs}$} \\ \cline{3-5} \cline{7-9} \cline{11-13} \cline{15-17} \cline{19-21}
$|V|$                                            & N & $\mu_t$ (s)      & $\sigma_t$  (s)   & Gap  (\%) &  & $\mu_t$ (s)      & $\sigma_t$  (s)   & Gap  (\%)    &  & $\mu_t$ (s)      & $\sigma_t$  (s)   & Gap  (\%)  & & $\mu_t$ (s)      & $\sigma_t$  (s)   & Gap  (\%)    & & $\mu_t$ (s)      & $\sigma_t$  (s)   & Gap  (\%)  \\ \hline
7                                                     & 100        & 6.3                   & 3.8                     & -               & & 1.4                   & 1.1                     & -              & & 4.9                   & 3.2                      & -              &  & 1.2                   & 0.5                      & -               & & 1.1                    & 0.5                       & -                 \\
                                                      & 200        & 10.2                  & 6.4                     & -               & & 2.0                   & 1.1                     & -              &  & 7.9                   & 5.7                      & -              &  & 1.9                   & 0.9                      & -              &  & 1.6                    & 0.6                       & -                 \\
                                                      & 500        & 15.0                  & 8.8                     & -               & & 3.2                   & 1.8                     & -              &  & 10.2                  & 6.5                      & -              &  & 2.4                   & 1.0                      & -              &  & 2.1                    & 0.8                       & -                 \\
                                                      & 1000       & 19.5                  & 11.2                    & -               & & 4.4                   & 2.5                     & -              &  & 11.9                  & 8.0                      & -              & & 2.8                   & 1.1                      & -               & & 2.4                    & 0.9                       & -                 \\ \hdashline
8                                                     & 100        & 29.3                  & 19.7                    & -              &  & 11.4                  & 10.5                    & -              &  & 51.7                  & 59.4                     & -              &  & 9.4                   & 6.6                      & -               & & 5.2                    & 4.4                       & -                 \\
                                                      & 200        & 50.5                  & 31.5                    & -              &  & 19.1                  & 16.5                    & -              &  & 83.2                  & 83.4                     & -              &  & 16.7                  & 22.9                     & -               & & 8.1                    & 6.9                       & -                 \\
                                                      & 500        & 88.9                  & 53.3                    & -               & & 32.3                  & 24.9                    & -              &  & 145.0                 & 146.2                    & -              &  & 24.9                  & 26.4                     & -               & & 13.4                   & 10.9                      & -                 \\
                                                      & 1000       & 127.9                 & 81.9                    & -               & & 44.3                  & 35.9                    & -              &  & 159.1                 & 150.8                    & (1)  0.31      &  & 30.7                  & 26.5                     & -              &  & 15.6                   & 14.2                      & -                 \\ \hdashline
9                                                     & 100        & 170.3                 & 157.6                   & (2)  0.33      &  & 35.3                  & 27.3                    & -              &  & 187.5                 & 130.3                    & (9)  0.54      &  & 43.1                  & 39.0                     & -              &  & 25.0                   & 19.1                      & -                 \\
                                                      & 200        & 238.9                 & 165.1                   & (7)  0.34       & & 68.4                  & 53.1                    & -               & & 315.6                 & 170.2                    & (14) 0.54      &  & 80.0                  & 96.4                     & -              &  & 41.1                   & 49.4                      & -                 \\
                                                      & 500        & 263.6                 & 140.3                   & (12) 0.38       & & 122.6                 & 120.3                   & (1)  0.20       & & 357.0                 & 170.0                    & (17) 0.55      &  & 105.0                 & 79.9                    & (1)  0.16       & & 58.6                   & 49.1                      & -                 \\
                                                      & 1000       & 317.7                 & 140.4                   & (16) 0.44       & & 204.2                 & 153.5                   & (1)  0.13      &  & 366.3                 & 198.4                    & (22) 0.55       & & 155.6                 & 100.9                    & (1)  0.07       & & 87.1                   & 66.5                      & -                 \\ \hdashline
10                                                    & 100        & 279.1                 & 159.7                   & (12) 0.28      &  & 91.8                  & 61.7                    & -              &  & 486.5                 & 253.8                      & (25) 0.61       & & 169.3                 & 137.1                    & (3)  0.16       & & 133.7                  & 152.2                     & -                 \\
                                                      & 200        & 258.5                 & 161.2                   & (23) 0.34      &  & 160.9                 & 117.9                   & -               & & 344.2                 & 361.7                      & (28) 0.61      &  & 238.5                 & 179.6                    & (4)  0.18      &  & 158.8                  & 138.3                     & (2)  0.16         \\
                                                      & 500        & 565.2                 & 58.4                    & (26) 0.48      &  & 245.3                 & 151.4                   & (6)  0.13       & & 283.9                 & 0.0                      & (29) 0.69       & & 264.4                 & 193.9                    & (12) 0.18       & & 223.5                  & 170.9                     & (7)  0.16         \\
                                                      & 1000       & 479.0                 & 89.4                    & (28)  0.53      & & 294.1                 & 149.2                  & (13) 0.18       & & 191.0                 & 0.0                      & (29) 0.72      &  & 266.0                 & 170.8                    & (18) 0.26       & & 273.6                  & 189.6                     & (8)  0.19        
\end{tabular}%
}
\caption{Computational performance of the DEF and L-shaped algorithms for the SAA problem of small-sized instances}
\label{tbl: mms-exact-solution-results table}
\end{table}

Observe from Table \ref{tbl: mms-exact-solution-results table} that using the improved model instead of the standard model decreased the computational time and optimality gap of both the DEF and L-shaped algorithm drastically. In particular, the solution time decreased for all instances. On average (over different $|V|$ and $N$), we observe a 67\% and 70\% decrease for the DEF and L-shaped algorithm, respectively. Additionally, the decrease in the standard deviation is around 64\% and 74\% for the DEF and L-shaped algorithm, respectively, when non-optimal solutions are left out. Moreover, the number of instances that could not be solved optimally is reduced by using the improved model: on average, 83\% and 78\% of those instances are solved optimally with $D_{imp}$ and $L_{imp}$, respectively. Additionally, the remaining non-optimal solutions are enhanced as a reduction in optimality gaps is achieved. 

Another drastic improvement is provided by our cut selection strategy. Comparing $L_{imp}$ and $L_{imp-cs}$ in Table \ref{tbl: mms-exact-solution-results table} shows that the mean and standard deviation of the computational time, and optimality gap are reduced by 35\%, 33\%, and 18\%, on average. Furthermore, an optimal solution is found by $L_{imp-cs}$ for 56\% of the instances that could not be solved optimally within the time limit by  $L_{imp}$. 

Finally, we compare $D_{imp}$ and $L_{imp-cs}$. 
Observe from Table \ref{tbl: mms-exact-solution-results table} that 
$L_{imp-cs}$ resulted in lower mean computational time by 31\%, 59\%, 45\%, and 1\% for instances with 7, 8, 9, and 10 vehicles, respectively, and by 15\%, 28\%, 38\%, and 45\% for instances with 100, 200, 500, 1000 scenarios, respectively. In the same order, the variance decreased by 56\%, 58\%, 38\%, and -52\% for instances with 7,8,9, and 10 vehicles, respectively, and by -1\%, 18\%, 41\%, and 42\% for instances with 100, 200, 500, 1000 scenarios, respectively. We conclude that our L-shaped algorithm outperforms the DEF  for instances with up to 9 vehicles, and they provide comparable results for instances with 10 vehicles. Additionally, the superiority of the L-shaped algorithm over the DEF escalates as the number of scenarios increases. 

\subsection{Heuristic Solution Approach}\label{section: mms-heuristic-results}
In this section, we present results for the solution quality and computational performance of the TS algorithm. The solution quality is evaluated by employing the MRP scheme, explained in Section \ref{mms: solution quality assessment}. We also assess the computational performance of the TS algorithm in various aspects.

We set the operator selection probabilities (weights) based on our preliminary experiments. The weights of swap, forward insertion, backward insertion, and inversion are 0.45, 0.10, 0.15, 0.30, respectively. We set the time limit for the one-scenario problem to 10 seconds, i.e., $\tau_{one}=10$ seconds, which leaves $\tau_{full}=590$ seconds. Finally, based on the results of the quality assessment, we set the sample size $N$ to 1000 for the computational performance assessments.

\subsubsection{Solution Quality}
\begin{figure}[h]
\captionsetup{justification=centering}
    \centerline{\includegraphics[scale=0.7]{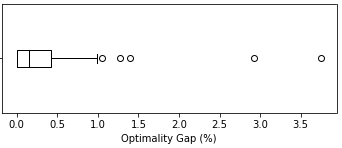}}
    \caption{Solution quality of the SAA problem}
    \label{fgr:ts_solution_quality}
\end{figure}

We solved the SAA problem of large-sized instances  using the TS algorithm. To assess the solution quality, we then used the proposed MRP integrated SAA scheme, given in Algorithm \ref{pseudo-code:MRP-integrated-SAA}, with the replication $M=30$, MRP sample size $N=20000$, $\alpha=0.05$ (95\% confidence interval), $\epsilon=0.05$, and  the list of sample sizes $N_{list}=\{1000, 2500, 5000\}$. Table \ref{tbl: mms-ts-simulated-performance} reports the  key performance result to show the performance of the MRP integrated SAA scheme. While the maximum optimality gap is 3.7\%, the average optimality gap is 0.28\% which indicates that solving the SAA problem with the proposed TS algorithm can produce high-quality solutions. Figure \ref{fgr:ts_solution_quality} further shows that the optimality gap for most of the solutions is less than 1\% with only five outliers out of 90 instances. 

\begin{table}[h]
\centering
\scalebox{0.8}{
\begin{tabular}{lllll}
\cellcolor[HTML]{FFFFFF}\begin{tabular}[c]{@{}l@{}}Statistical   Lower \\      Bound $\mathbb{E}[z_{N}]$ \end{tabular} & \cellcolor[HTML]{FFFFFF}\begin{tabular}[c]{@{}l@{}}Estimated Objective\\      Value $\mathbb{E}[Q(\hat{\textit{x}}_{N}, \xi_{\omega})]$ \end{tabular} & \cellcolor[HTML]{FFFFFF}\begin{tabular}[c]{@{}l@{}}Optimality \\      Gap $\Bar{G}_{N}$\end{tabular} &  &  \\ \cline{1-3}
\cellcolor[HTML]{FFFFFF}239.59    & \cellcolor[HTML]{FFFFFF}240.26       & \cellcolor[HTML]{FFFFFF}0.67 (0.28\%)                             
\end{tabular}}
\caption{Solution quality of the MRP integrated SAA} 
\label{tbl: mms-ts-simulated-performance}
\end{table}

\begin{figure}[h]
\captionsetup{justification=centering}
    \centerline{\includegraphics[scale=0.7]{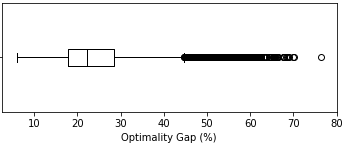}}
    \caption{Solution quality of one-scenario problem}
    \label{fgr:ts_det_quality}
\end{figure}

Moreover, we assess the solution quality of the one-scenario problem over large-sized instances by using the same procedure in order to show its robustness. Observe from Figure \ref{fgr:ts_det_quality} that the average optimality gap is 24.8\%, the maximum optimality gap is 76.5\%, and the standard deviation is 10.8\%. Comparing the performance of the SAA and one-scenario problems demonstrates that we can generate robust solutions by considering vehicle failures. Accordingly, we reassure with the industry-sized instances that we can reduce the work overloads by more than 20\% by considering stochastic car failures and solving the corresponding problem efficiently. 

\subsubsection{Computational Performance}
\label{sec: compu}
To assess the computational performance of the TS algorithm, we conducted different tests: 1) compared the solution found with the TS algorithm with the solution found by an off-the-shelf solver for the one-scenario problem of medium-sized instances, 2) compared the solution found with the TS algorithm  with the optimal solution found by $L_{imp-cs}$ approach for the SAA problem of small-sized instances, 3) compared the solution found with the TS algorithm  with that of a simulated annealing (SA) algorithm for the SAA problem of large-sized instances, 4) and  analyzed the convergence of TS algorithm for the one-scenario and SAA problems of large-sized instance. We executed 30 runs for each of the instances and tests. 

Table \ref{tbl: mms-ts-vs-gurobi-det} reports the results for the first set of computational experiments for the one-scenario problem. We note that we generated 30 instances that could be solved within three hours time limit with Gurobi (solving the DEF). The minimum, average, maximum, and standard deviation of the computational time (in seconds) are shown for the  TS algorithm and Gurobi. Additionally, the average number of movements for the TS algorithm is reported in order to provide some insight about the efficiency of the implementation. The optimal solutions, for all 30 instances and for all 30 runs, are found in under 10 seconds by the TS algorithm. The average computational times are 1140 and 0.33 seconds for the Gurobi solver and TS algorithm, respectively. These results show that the proposed TS algorithm can consistently provide optimal solutions to the one-scenario problem in a very short amount of time by avoiding any local optima.

\begin{table}[h]
\centering
\scalebox{0.8}{
\begin{tabular}{
>{\columncolor[HTML]{FFFFFF}}l 
>{\columncolor[HTML]{FFFFFF}}c 
>{\columncolor[HTML]{FFFFFF}}c 
>{\columncolor[HTML]{FFFFFF}}c 
>{\columncolor[HTML]{FFFFFF}}c 
>{\columncolor[HTML]{FFFFFF}}c 
>{\columncolor[HTML]{FFFFFF}}c }
       & \multicolumn{4}{c}{\cellcolor[HTML]{FFFFFF}Time (s)}                                                                                                                                                 & \multicolumn{1}{l}{\cellcolor[HTML]{FFFFFF}} & \multicolumn{1}{l}{\cellcolor[HTML]{FFFFFF}Move (\#)} \\ \cline{2-5} \cline{7-7} 
Method & \multicolumn{1}{c}{\cellcolor[HTML]{FFFFFF}Min} & \multicolumn{1}{c}{\cellcolor[HTML]{FFFFFF}Mean} & \multicolumn{1}{c}{\cellcolor[HTML]{FFFFFF}Max} & \multicolumn{1}{c}{\cellcolor[HTML]{FFFFFF}Std. Dev.} &                                              & Mean                                                            \\ \hline
Gurobi & 2.37                                               & 1140.91                                            & 9373.08                                            & 2613.95                                            &                                              & -                                                              \\
TS     & 6e-4                                           & 0.33                                               & 9.44                                               & 0.81                                               &                                              & 16940                                                         
\end{tabular}}
\caption{Computational performance of Gurobi and TS for the one-scenario problem of medium-sized instances}
\label{tbl: mms-ts-vs-gurobi-det}
\end{table}


\begin{table}[h]
\centering
\scalebox{0.8}{
\begin{tabular}{
>{\columncolor[HTML]{FFFFFF}}l l
>{\columncolor[HTML]{FFFFFF}}l 
>{\columncolor[HTML]{FFFFFF}}l 
>{\columncolor[HTML]{FFFFFF}}l 
>{\columncolor[HTML]{FFFFFF}}l 
>{\columncolor[HTML]{FFFFFF}}l 
>{\columncolor[HTML]{FFFFFF}}l 
>{\columncolor[HTML]{FFFFFF}}l 
>{\columncolor[HTML]{FFFFFF}}l 
>{\columncolor[HTML]{FFFFFF}}c 
>{\columncolor[HTML]{FFFFFF}}c 
>{\columncolor[HTML]{FFFFFF}}c }
                & \multicolumn{4}{c}{\cellcolor[HTML]{FFFFFF}Time (s)}                             &  & \multicolumn{3}{c}{\cellcolor[HTML]{FFFFFF}Optimality Gap (\%)}      & \multicolumn{1}{c}{\cellcolor[HTML]{FFFFFF}} & \begin{tabular}[c]{@{}c@{}}Optimally \\      Solved (\%)\end{tabular} & \multicolumn{1}{c}{\cellcolor[HTML]{FFFFFF}\textbf{}} & \begin{tabular}[c]{@{}c@{}}Move \\  (\#)\end{tabular} \\ \cline{2-5} \cline{7-9} \cline{11-11} \cline{13-13}
Method & \cellcolor[HTML]{FFFFFF}Min & Mean & Max & Std. Dev. &           & Mean & Max & Std. Dev. &                                            & \multicolumn{1}{l}{\cellcolor[HTML]{FFFFFF}}                          &                                            & Mean       \\ \hline
Gurobi          & 5.02                                    & 68.91           & 210.06          & 50.99           &           & 0                     & 0                      & 0                 &                                                       & 100                                                                            &                                                       & -                  \\
TS              & \cellcolor[HTML]{FFFFFF}1e-4          & 0.08            & 0.28            & 0.08            &           & 0.14                  & 2.79                   & 0.41              &                                                       & 83                                                                             &                                                       & 628               
\end{tabular}
}
\caption{Computational performance of Gurobi and TS for the SAA problem of smalls-sized instances}
\label{tbl: mms-ts-vs-gurobi-stoc}
\end{table}

Recall that as demonstrated in Section \ref{sec: exact_comp_per}, $L_{imp-cs}$ outperformed Gurobi in solving the SAA problem for small-sized instances. In  the second set of experiments, we compared the computational effectiveness of solving the SAA problem of small-sized instances with the TS algorithm and $L_{imp-cs}$. To this end, we chose a sample size of 1000 and solved 30 small-sized instances optimally by using $L_{imp-cs}$. We then solved the same set of instances using the TS algorithm until either an optimal solution was obtained or the time limit was reached. Table \ref{tbl: mms-ts-vs-gurobi-stoc}  shows that TS algorithm was able to find optimal solutions in 83\% of the experiments, 746 out of 900. However, the average optimality gap for these non-optimal solutions is 0.14\%, indicating that the TS algorithm is reliable in terms of optimality. For the TS algorithm, we recorded the computational time as the time until the last improvement is done, so we can see that the TS algorithm is very efficient, with an average runtime of 0.08 seconds. 

We observed that the TS algorithm terminated at a local optima often when the number of vehicles is very small.
Our hypothesis is that there are plateaus with the same objective value when the number of vehicles is large. However,  this is not the case when there are only 10 vehicles as there are a limited number of sequences and each sequence has a different objective function value. Hence, in the third set of experiments, we compared the TS algorithm and a SA algorithm for large-sized instances to analyze the impact of the tabu list and accepting worse solutions on escaping a local optima. We utilized the same local search algorithm for the SA with two differences: 1) disabled the tabu rules and 2) enabled accepting worse solutions based on the acceptance criterion. For the SA algorithm, we set the starting temperature $T_{init}=10$. We also adopted a geometric cooling with the cooling constant $\alpha=0.999$. The acceptance ratio is calculated using the Boltzmann distribution $P(\Delta f, T) = e^{-\frac{f(x')-f(x)}{T}}$, where $x'$ is a new solution, $x$ is the incumbent solution, and $T$ is the current temperature.

\begin{figure}[h]
\captionsetup{justification=centering}
    \centerline{\includegraphics[scale=0.5]{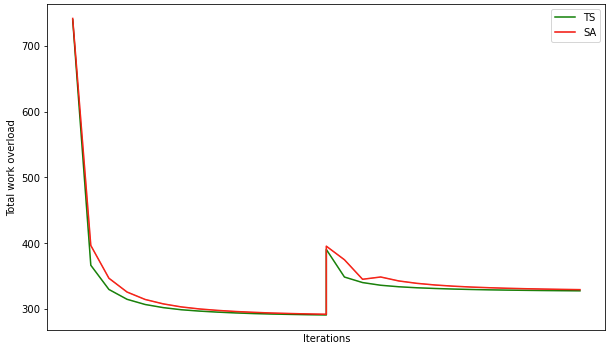}}
    \caption{Convergence comparison of the TS and SA algorithms}
    \label{fgr:mms-ts-vs-sa}
\end{figure}

We note that the Kruskal-Wallis test shows no significant difference between the computational performance of the TS and SA algorithms at a 95\% confidence level. However, as illustrated in Figure \ref{fgr:mms-ts-vs-sa}, the TS algorithm produces better results and converges faster than the  SA algorithm (averaged over  of 900 runs). This result shows that the proposed TS algorithm is capable of exploiting the search space while generally avoiding premature convergence to local optima. Accordingly, we conclude that there is no need to accept worse solutions in the local search. 

\begin{figure}[h]
\captionsetup{justification=centering}
    \centerline{\includegraphics[scale=0.7]{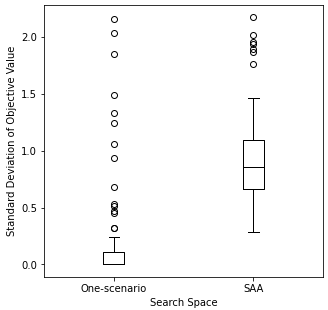}}
    \caption{Convergence of the objective value with TS algorithm for the one-scenario and SAA problems}
    \label{fgr:ts-convergence-det-stoc}
\end{figure}

Finally, in the last set of experiments, we conducted an analysis in order to provide insight into the reliability of the proposed TS algorithm's convergence. In particular, in Figure \ref{fgr:ts-convergence-det-stoc}, we presented the box plots of  the standard deviation of the objective values for the one-scenario and SAA problems of large-sized instances. Each data point represents the standard deviation of 30 runs (for each of the 90 large-sized instances). The average standard deviations for the one-scenario and SAA  problems are 0.19 and 0.93, while the  means of the objective values are 212.77 and 239.69, respectively. Accordingly, the average coefficient of variations, the ratio between the average standard deviation and mean, are 9e-4 and 4e-3, which indicates that the proposed TS algorithm provides highly reliable results for both of the problems. 

\section{Conclusion}\label{section: mms-conclusion}
This paper studied mixed-model sequencing (MMS) problem with stochastic failures. To the best of our knowledge, this is the first study that considers stochastic failures of products in MMS. The products (vehicles) fail according to various characteristics and are then removed from the sequence, moving succeeding vehicles forward to close the gap. Vehicle failure may cause extra work overloads that could be prevented by generating a robust sequence at the beginning. 
Accordingly, we formulated the problem as a two-stage  stochastic program, and improvements were presented for the second-stage problem. We employed the sample average approximation approach to tackle the exponential number of scenarios. We developed L-shaped decomposition-based algorithms to solve small-sized instances. 
The numerical experiments showed that the L-shaped algorithm outperforms  the deterministic equivalent formulation, solved with an off-the-shelf solver, in terms of both solution quality and computational time. 
To solve industry-sized instances efficiently, we developed a greedy heuristic and a tabu search algorithm that is accelerated with problem-specific tabu rules. Numerical results showed that we can provide good quality solutions, with less than a 5\% statistical optimality gap, to industry-sized instances in under ten minutes.
The numerical experiments also indicated that  we can generate good quality robust solutions by utilizing a sample of scenarios. In particular, we can reduce the work overload by more than 20\%, for both small- and large-sized instances, by considering possible car failures. 

\subsection{Managerial Insights}
Car manufacturers are facing several challenges due to the increasing ratio of EVs in production. EVs have significant differences compared to non-EVs which require specific treatment when creating the sequence for the mixed-model assembly line. One of the main challenges is the battery installation process. Unlike traditional vehicles, EVs have large and heavy batteries that need to be installed in a specific order to avoid damaging the vehicle or the battery itself. Accordingly, a huge processing time variation at the battery loading station arises from this difference, which requires special treatment to ensure that the assembly line can continue to produce high-quality vehicles efficiently.

We have observed that consecutive EVs induce a significant amount of work overload, which generally requires line stoppage even with the help of utility workers. Planning the sequence by ruling out back-to-back EVs does not guarantee that there will not be any occurrence of consecutive EVs. The failure of vehicles disrupts the planned sequence, and the necessity of considering car failures during the planning process increases as the difference between product types expands. 

In this study, we focused on generating robust schedules that take into account the possible deleterious effects resulting from the divergence between electric and non-electric vehicles. However, it is worth noting that our proposed solution methodologies are equally applicable to any feature that causes similar variations at specific work stations.

\subsection{Future Research}
One direction for future research is the reinsertion of the failed vehicles back into the sequence. Even though the reinsertion process is being conducted via a real-time decision, a robust approach may increase the efficiency of production. Another direction for future research is to include stochastic processing times in addition to stochastic product failures. This may potentially generate more robust schedules, particularly in case a connection between failures and processing times is observed. Finally, there are similarities between MMS and some variants of the traveling salesman problem (TSP). Since the TSP is one of the most studied combinatorial optimization problems, the state-of-art solution methodologies presented for TSP may be adapted to MMS.

\section*{Acknowledgements}
The authors acknowledge Clemson University for generous allotment of compute time on Palmetto cluster.

\appendix
\section{Tabu List for Local Search Algorithm} \label{appendix: mms-tabu-list}
Assume that we have two positions is selected for any operator to be applied, $t_1, t_2$ with $ t_1<t_2$. The tabu movements for each operator is described below under two cases: the vehicle at the position $t_1$ is 1) an EV, 2) not an EV. \\
\textbf{\underline{\textit{Swap}}} \\
1) The position $t_2$ cannot be a neighbor of an EV, e.g., the vehicles at the positions $t_{2}-1$ and $t_{2}+1$ cannot be an EV. \\
2) The vehicle at the position $t_2$ cannot be an EV if the position $t_1$ is a neighbor of an EV. \\
\textbf{\underline{\textit{Forward Insertion}}} \\
1) The vehicles at the positions $t_2$ or $t_{2}-1$ cannot be an EV. \\
2) At most one of the vehicles at the positions $t_{1}-1$ and $t_{1}+1$ can be an EV. \\
\textbf{\underline{\textit{Backward Insertion}}} \\
1) The vehicles at the positions $t_1$ or $t_{1}-1$ cannot be an EV. \\
2) At most one of the vehicles at the positions $t_{2}-1$ and $t_{2}+1$ can be an EV. \\
\textbf{\underline{\textit{Inversion}}} \\
1) The position $t_2$ cannot be a left neighbor of an EV, e.g., the vehicle at the position $t_{2}+1$ cannot be an EV. \\
2) If the vehicle at the position $t_1$ is a right neighbor of an EV, then the vehicle at the position $t_2$ cannot be an EV. 

\begin{spacing}{0.0}
\bibliographystyle{elsarticle-harv} 
\bibliography{cas-refs}
\end{spacing}

\end{document}